\documentclass{cernrep}
\usepackage{axodraw} 
\usepackage{rotating}
\usepackage{epic}
\usepackage{curves}
\def\dotprod{\mathord{\cdot}}
\def\btorho{\bar B^0\to\rho^+ l^- \bar\nu_l}
\def\btopi{\bar B^0\to\pi^+ l^- \bar\nu_l}
\def\btokstargamma{\bar B\to K^* \gamma}
\def\gtrsim{\mathrel{\lower .7ex\hbox{$\buildrel\textstyle>\over\sim$}}}
\def\lesssim{\mathrel{\lower .7ex\hbox{$\buildrel\textstyle<\over\sim$}}}
\newcommand{\lqcd}{\Lambda_{\mathrm{QCD}}}
\def\kev{\,\mathrm{ke\kern-0.1em V}}
\def\mev{\,\mathrm{Me\kern-0.1em V}}
\def\gev{\,\mathrm{Ge\kern-0.1em V}}
\def\tev{\,\mathrm{Te\kern-0.1em V}}
\def\etal{et al.}
\def\qsqmax{q^2_{\mathrm{max}}}
\newcommand{\kzero}{|K^0\rangle}
\newcommand{\kone}{|K_1\rangle}
\newcommand{\ktwo}{|K_2\rangle}
\newcommand{\ks}{|K_S\rangle}
\newcommand{\kl}{|K_L\rangle}
\newcommand{\kzerobar}{|\bar K^0\rangle}
\newdimen\unit
\def\point#1 #2 #3{\vbox to0pt{\kern-#2\unit
  \hbox{\kern#1\unit$#3$}\vss}
 \nointerlineskip}

\newcommand{\err}[2]{%
{{\renewcommand{\arraystretch}{0.4}%
\ensuremath{\mathop{\raisebox{0.1\height}{\scriptsize
$\begin{array}{@{}c@{}}+\\-\end{array}$}}%
\raisebox{0.1\height}{\scriptsize
$\begin{array}{@{}r@{}}#1\\#2\end{array}$}}}}%
}

\def\dotprod{\mathord{\cdot}}
\def\vec#1{\mathbf{#1}}

\begin{document}
\begin{flushright}SHEP 98/01\\
hep-ph/9801343
\end{flushright}
\vspace{3em}

\begin{center}
{\Large\bfseries Flavour Physics}\\[2em]
C.T.~Sachrajda\\[1em]
Department of Physics and Astronomy, University of Southampton\\
Southampton SO17 1BJ, UK
\end{center}
\vfill

\begin{center}\textbf{Abstract}\end{center}
\begin{quote}
In these four lectures I review the theory and phenomenology of weak decays
of quarks, and their r\^ole in the determination of the parameters of the
Standard Model of particle physics, in testing subtle features of the theory
and in searching for signatures of \emph{new} physics. Attempts to understand
$CP$-violation in current and future experiments is discussed.
\end{quote}
\vfill

\begin{center}
To appear in the proceedings of the 1997 European School of High-Energy
Physics, Menstrup, Denmark, 15~May - 7~June 1997. 
\end{center}
\vfill

\begin{flushleft}
January 1998
\end{flushleft}
\newpage
\setcounter{page}{1}
\title{FLAVOUR PHYSICS}
\author{C. T. SACHRAJDA}

\institute{Department of Physics and Astronomy,\\
University of Southampton,\\
Southampton SO17 1BJ,\\
England}

\maketitle

\begin{abstract}
In these four lectures I review the theory and phenomenology of weak decays
of quarks, and their r\^ole in the determination of the parameters of the
Standard Model of particle physics, in testing subtle features of the theory
and in searching for signatures of \emph{new} physics. Attempts to understand
$CP$-violation in current and future experiments is discussed.
\end{abstract}

\section{Lecture 1: Introduction}

\emph{Flavourdynamics}, the study off the electroweak Lagrangian and its
implications, is one of the central areas of research in particle
physics.  For example, in the United Kingdom the \emph{Mission
Statement} of the Particle Physics Community, as defined by our
research council, is to obtain insights into the following three
fundamental questions:
\begin{itemize}
\item the origin of mass,
\item the 3 generations of elementary particles and their weak asymmetries,
\item[] \hspace{0.5 in} and
\item the nature of dark matter. 
\end{itemize}
The second of these topics is flavourdynamics, which is the subject of
this lecture course. I will review the theoretical and experimental
work being done in an attempt to determine the parameters of the
Standard Model (SM) accurately, to test subtle properties of the SM
and to define and search for signatures of \emph{New
Physics}. Although this primarily involves weak interactions, the
major theoretical difficulty in interpreting experimental data on weak
hadronic processes is controlling the non-perturbative strong
interaction effects necessarily present in these decays. For this
reason any review of flavour physics must contain a discussion also of
these QCD effects, and much of the presentation below concerns this
subject.

In the next few years much of the effort of the experimental community
will be concerned with trying to gain an understanding of
$CP$-violation. Indeed, one of the Sakharov conditions for generating
a matter-antimatter asymmetry in the universe in the big-bang
cosmology is the requirement for the existence of $CP$-violation. The
existence of three generations of quarks and leptons, and hence of a
complex phase in the mixing matrix, implies the presence of
$CP$-violation at some level. Very little is known, however, about the
value of this phase and hence of the magnitude of $CP$-violation
induced by this mechanism, and also about other possible sources. It
is hoped that the intensive studies, which are about to begin will
significantly further our understanding and this will be one of the
main topics of discussion towards the end of this lecture course.

The program for the four lectures is as follows:
\begin{enumerate}
\item[1:] The first lecture will be an introduction to
weak decays in the Standard Model. Among the topics which will be
covered will be charged and neutral currents, the
Cabibbo-Kobayashi-Maskawa matrix and unitarity triangles, parity and
charge conjugation symmetries, effective Hamiltonians and Operator
Product Expansions and the Heavy Quark Effective Theory.
\item[2:] In the second lecture I will review the
determination of the $V_{cb}$ and $V_{ub}$ matrix elements from
semileptonic decays of $B$-mesons. Lattice computations, which
provide the opportunity to evaluate non-perturbative strong interaction
effects in weak decays in general, are introduced in this lecture.
\item[3:] The third lecture will contain a review of $K^0$-$\bar K^0$
and $B^0$-$\bar B^0$ mixing.
\item[4:] Finally, in the last lecture I review $CP$-violation in
$B$-decays and the subject of inclusive $B$-decays.
\end{enumerate}

The theoretical framework introduced in these lectures will be applied
to several important physical processes. It will not be possible,
however, to discuss the full range of interesting processes which are
providing, or will provide, fundamental physical information. A much
larger set of physical quantities is considered in detail in the
beautiful review by Buras and Fleischer~\cite{bf}, to which I refer
the student, and from which I will quote extensively. I also refer the
reader to ref.~\cite{dgh} for a broad introduction to the
\emph{Dynamics of the Standard Model} and to references~\cite{stone}
and \cite{jarlskog} for modern reviews of \emph{$B$-decays} and
\emph{$CP$-violation} respectively. I will assume a familiarity with
elementary Quantum Field Theory (as discussed for example by
J.~Petersen and V.~Zakharov at this school~\cite{zakharov}), but will
try to avoid technical complexities, focussing instead on the
underlying ideas.

\subsection{The Interactions of Quarks and Gauge Bosons}
\label{subsec:interactions}

\begin{figure}
\begin{center}
\begin{picture}(360,90)(0,15)
\ArrowLine(0,100)(30,100)\ArrowLine(30,100)(60,100)
\Text(15,107)[b]{$i$}\Text(45,105)[b]{$j$}
\ArrowLine(100,100)(130,100)\ArrowLine(130,100)(160,100)
\Text(115,107)[b]{$i$}\Text(145,107)[b]{$i$}
\ArrowLine(200,100)(230,100)\ArrowLine(230,100)(260,100)
\Text(215,107)[b]{$i$}\Text(245,107)[b]{$i$}
\ArrowLine(300,100)(330,100)\ArrowLine(330,100)(360,100)
\Text(315,107)[b]{$i$}\Text(345,107)[b]{$i$}
\Gluon(30,100)(30,60){2.5}{5}\Text(32,55)[t]{$W^\pm$}
\Text(30,30)[t]{(a)}
\Gluon(130,100)(130,60){2.5}{5}\Text(132,55)[t]{$Z^0$}
\Text(130,30)[t]{(b)}
\Gluon(230,100)(230,60){2.5}{5}\Text(230,53)[t]{$\gamma$}
\Text(230,30)[t]{(c)}
\Photon(330,100)(330,60){2}{7}\Text(330,55)[t]{${\mathcal G}$}
\Text(330,30)[t]{(d)} \end{picture} \caption{The basic vertices
representing the interactions of the quarks with the gauge bosons. The
labels $i$ and $j$ represent the flavour quantum number
($i,j=u,d,c,s,t,b$).} \label{fig:vertices}\end{center} \end{figure}
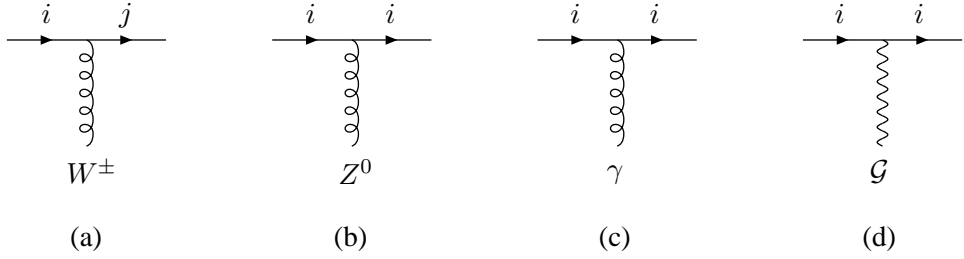

The interactions between the quarks and gauge bosons in the Standard
Model are illustrated in Fig.~\ref{fig:vertices}. In these lectures we
will be particularly interested in the weak interactions. The vertex for
the charged current interaction in which quark flavour $i$ changes to
$j$ is depicted in Fig~\ref{fig:vertices}(a) and has the Feynman rule
\begin{equation}
i\frac{g_2}{2\sqrt 2} V_{ij} \gamma_\mu(1-\gamma_5)\ ,
\label{eq:ccvertex}\end{equation}
where $g_2$ is the coupling constant of the $SU(2)_L$ gauge group and
$V_{ij}$ is the $ij$ element of the Cabibbo-Kobayashi-Maskawa (CKM)
matrix ($V_{ji}=V_{ij}^*$)~\cite{ckm}. Eq.~(\ref{eq:ccvertex})
illustrates the $V{-}A$ (vector${-}$axial-vector) structure of the
charged-current interactions.

\begin{figure}
\begin{center}
\begin{picture}(170,40)(0,60)
\Line(10,100)(70,100)\Line(10,60)(70,60)
\Gluon(40,100)(40,60){2.5}{5}
\Text(30,80)[r]{$W$}\Line(85,80)(100,80)\ArrowLine(100,80)(102,80)
\Line(147,80)(132,93)\Line(147,80)(162,93)
\Line(147,80)(132,67)\Line(147,80)(162,67)
\end{picture}
\end{center}
\caption{Approximation of the $W$-exchange interaction, by the
four-fermion current-current vertex.\label{fig:fermi}}
\end{figure}
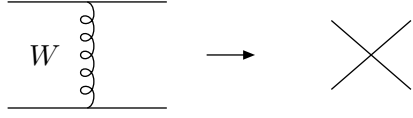

At low energies, so that the momentum in the $W$-boson is much smaller
than its mass $M_W$, the four-fermion interaction mediated by the
$W$-boson can be approximated by the local Fermi $\beta$-decay
interaction with coupling $G_F$, where
\begin{equation}
\frac{G_F}{\sqrt 2} = \frac{g_2^2}{8 M_W^2}\ ,
\label{eq:gfdef}\end{equation}
(see Fig.~\ref{fig:fermi}).

\subsection{CKM - Matrix}
\label{subsec:ckm}

The charged-current interactions are of the form
\begin{equation}
J_\mu^{CC} = (\bar u, \bar c, \bar t)_L\gamma_\mu V_{\mathrm{CKM}}
\left(\begin{array}{c}
d\\ s\\ b\end{array}\right)_{\!\!\! L}\ ,
\label{eq:cc}\end{equation}
where (assuming the Standard Model with 3 generations) the CKM-matrix,
$V_{\mathrm{CKM}}$ is a unitary 3$\times$ 3 one, relating the weak and
mass eigenstates. The 1996 particle data book~\cite{pdg} gives the
following values for the magnitudes of each of the elements:
\begin{equation}
\left(
\begin{array}{ccc}
0.9745 - 0.9757 & 0.219 - 0.224 & 0.002 - 0.05\\
0.218 - 0.224 & 0.9736 - 0.9750 & 0.036 - 0.046\\
0.004 - 0.014 & 0.034 - 0.046 & 0.9989 - 0.9993
\end{array}\right)\ .\label{eq:ckmvalues}\end{equation}
The subscript $L$ in eq.~(\ref{eq:cc}) represents {\em left handed},
($\psi_L=\frac{1}{2}(1-\gamma^5)\psi$).

If we have $2 N_g$ quark flavours, ($N_g$ is the number of generations)
then $V_{\mathrm{CKM}}$ is a
$N_g\times N_g$ unitary matrix. It therefore has $N_g^2$ real
parameters, however $(2N_g -1)$ of these can be absorbed into unphysical
phases of the quark fields~\footnote{The reason why the number of phases
is $(2 N_g -1)$ rather than $2N_g$, is that if the fields are multiplied
by the same phase factor then $J_\mu^{CC}$ is unchanged. Thus there is
one fewer phase, which can be absorbed.}, leaving $(N_g -1)^2$ physical
parameters to be determined.

In the two flavour case there is just one parameters, which is
conventionally chosen to be the Cabibbo angle:
\begin{equation}\left(
\begin{array}{rr}
\cos\theta_c & \ \sin\theta_c\\ 
-\sin\theta_c & \ \cos\theta_c  \end{array}\right)
\label{eq:v2}\end{equation}  

With three flavours there are 4 real parameters. Three of these can be
interpreted as angles of rotation in three dimensions (e.g. the three
Euler angles) and the fourth is a phase. The general parametrisation
recommended by the Particle Data Group~\cite{pdg} is
\begin{equation}\left(
\begin{array}{ccc}
c_{12}c_{13} & s_{12}c_{13} & s_{13}\exp(-i\delta_{13})\\
-s_{12}c_{23}-c_{12}s_{23}s_{13}\exp(i\delta_{13}) &
c_{12}c_{23}-s_{12}s_{23}s_{13}\exp(i\delta_{13}) & s_{23}c_{13}\\ 
s_{12}s_{23}-c_{12}c_{23}s_{13}\exp(i\delta_{13}) &
-c_{12}s_{23}-s_{12}c_{23}s_{13}\exp(i\delta_{13}) & c_{23}c_{13} 
\end{array}\right)\label{eq:ckmgenpar}\end{equation}
where $c_{ij}$ and $s_{ij}$ represent the cosines and sines respectively
of the three angles $\theta_{ij}$, $ij=12,\, 13\, 23$. $\delta_{13}$ is
the phase parameter. 

The parametrisation in eq.~(\ref{eq:ckmgenpar}) is general, but
awkward to use. For most practical purposes it is sufficient to
exploit the empirical fact that the elements get smaller as one moves
away from the diagonal of the matrix (see eq.~(\ref{eq:ckmvalues})\,),
and to use a simpler, but approximate parametrisation. It has become
conventional to use the Wolfenstein
parametrisation~\cite{wolfenstein}:
\begin{equation}
V_{\mathrm{CKM}}=\left(
\begin{array}{ccc}
1-\frac{\lambda^2}{2} & \lambda & A\lambda^3 (\rho - i \eta)\\
-\lambda & 1-\frac{\lambda^2}{2}\rule{0pt}{14pt} & A\lambda^2\\ 
A\lambda^3 (1- \rho - i \eta) & - A\lambda^2 \rule{0pt}{14pt}& 1
\end{array}\right)\ .
\label{eq:wolfenstein}\end{equation}
$\lambda$ is approximately the Cabibbo angle, and the terms which are
neglected in the Wolfenstein parametrisation are of $O(\lambda^4)$. Much
of the current effort of the particle physics community is devoted to
determining the four parameters $\lambda, A, \rho$ and $\eta$, with ever
increasing precision, and this will continue for some time to come. For
this reason I will spend a significant fraction of these lectures on the
theoretical issues related to the determination of these parameters.

\begin{figure}[th]
\begin{center}
\begin{picture}(250,125)(0,-20)
\Line(0,0)(200,0)\ArrowLine(0,0)(140,80)\ArrowLine(140,80)(200,0)
\Text(150,90)[b]{$A=(\bar\rho,\bar\eta)$}\Text(138,72)[t]{$\alpha$}
\Text(10,-10)[t]{$C=(0,0)$}\Text(22,2)[bl]{$\gamma$}
\Text(210,-10)[t]{$B=(1,0)$}\Text(188,2)[br]{$\beta$}
\Text(23,43)[l]{$\bar\rho + i \bar\eta$}
\Text(242,43)[r]{$1-(\bar\rho + i \bar\eta$)}
\end{picture}
\caption{Unitarity Triangle corresponding to the relation in
eq.~(\protect\ref{eq:ut}).}\label{fig:ut}
\end{center}\end{figure}
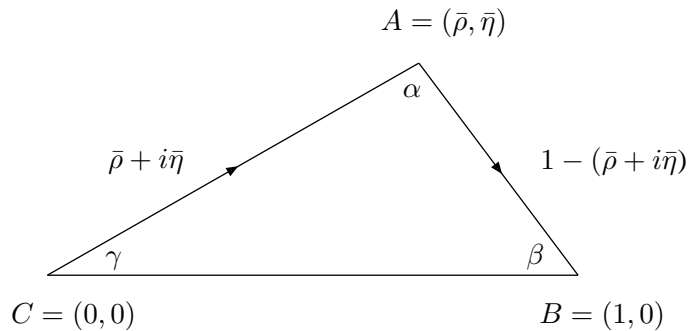

From the unitarity of the CKM-matrix we have a set of relations between
the entries. A particularly useful one is:
\begin{equation}
V_{ud}V_{ub}^* + V_{cd}V_{cb}^* + V_{td}V_{tb}^* =0 \ .
\label{eq:ut}\end{equation}
In terms of the Wolfenstein parameters, the components on the left-hand
side of eq.~(\ref{eq:ut}) are given by:
\begin{eqnarray}
V_{ud}V_{ub}^* & = & A\lambda^3[\bar\rho + i \bar\eta]+ O(\lambda^7)\nonumber\\
V_{cd}V_{cb}^* & = & -A\lambda^3 + O(\lambda^7)\label{eq:utelements}\\ 
V_{td}V_{tb}^* & = & A\lambda^3 [1 - (\bar\rho + i \bar\eta)] + O(\lambda^7)\nonumber\ ,
\end{eqnarray}
where $\bar\rho=\rho(1-\lambda^2/2)$ and $\bar\eta =
\eta(1-\lambda^2/2)$. The unitarity relation in eq.~(\ref{eq:ut}) can be
represented schematically by the famous ``unitarity triangle" of Fig.~\ref{fig:ut}
(obtained after scaling out a factor of $A\lambda^3$).

The notation in Fig.~\ref{fig:ut} is standard, and in the following
lectures I will describe current and future attempts to determine the
position of the vertex $A$, and hence the values of the angles
$\alpha,\beta$ and $\gamma$. By using many different processes to
overdetermine the position of $A$, one will be able to test the
consistency of the SM, and check for evidence of physics beyond this
model.

\subsubsection{The Cabibbo Sector}
\label{subsubsec:sinthetac}

Although much of the remainder of these lectures will be devoted to the
determination of the entries in the CKM-matrix, since the Cabibbo sector
is the best determined, I will not consider it beyond this subsection. I
briefly review the status of the four entries in the $2\times 2$
top-left submatrix of $V_{\mathrm{CKM}}$~\cite{pdg}.

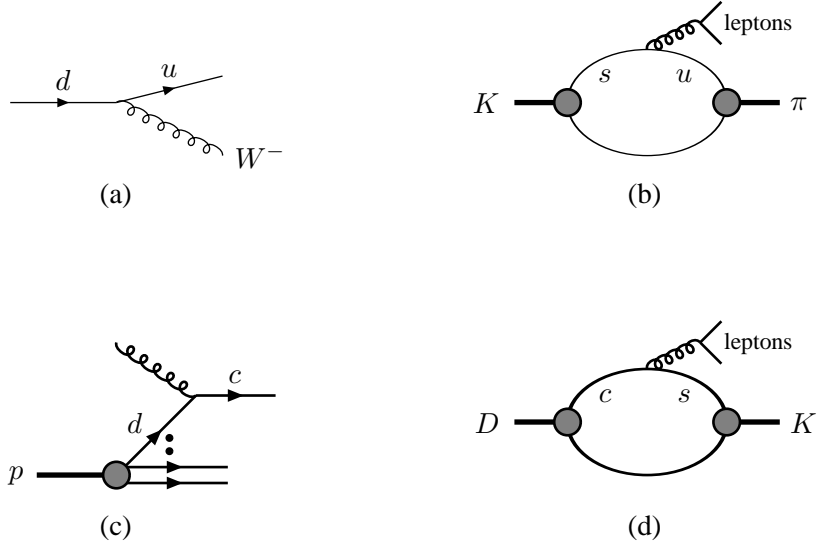
\begin{figure}[th]
\begin{center}
\begin{picture}(320,120)(0,20)
\ArrowLine(10,100)(50,100)\ArrowLine(50,100)(90,110)
\Gluon(50,100)(90,80){2}{6}
\Text(30,105)[b]{$d$}\Text(70,110)[b]{$u$}
\Text(95,80)[l]{$W^-$}\Text(50,70)[t]{(a)}
\Oval(250,100)(20,30)(0)
\SetWidth{2}\Line(220,100)(200,100)\Line(280,100)(300,100)
\SetWidth{1}
\GCirc(220,100){5}{0.5}\GCirc(280,100){5}{0.5}
\Gluon(250,120)(270,130){2}{4}
\Line(270,130)(278,138)\Line(270,130)(278,122)
\Text(280,130)[l]{{\footnotesize leptons}}
\Text(195,100)[r]{$K$}\Text(305,100)[l]{$\pi$}
\Text(235,108)[b]{$s$}\Text(265,108)[b]{$u$}
\Text(250,70)[t]{(b)}
\end{picture}
\begin{picture}(320,90)(0,50)
\SetWidth{2}\Line(20,80)(50,80)\SetWidth{1}
\Text(15,80)[r]{$p$}
\ArrowLine(50,80)(80,110)\Text(60,100)[r]{$d$}
\ArrowLine(80,110)(110,110)\Text(95,115)[b]{$c$}
\ArrowLine(52,77)(92,77)\ArrowLine(52,83)(92,83)
\Gluon(80,110)(50,130){2}{5}
\GCirc(50,80){5}{0.5}\GCirc(70,89){1}{0}\GCirc(70,94){1}{0}
\Text(50,65)[t]{(c)}
\Oval(250,100)(20,30)(0)
\SetWidth{2}\Line(220,100)(200,100)\Line(280,100)(300,100)
\SetWidth{1}
\GCirc(220,100){5}{0.5}\GCirc(280,100){5}{0.5}
\Gluon(250,120)(270,130){2}{4}
\Line(270,130)(278,138)\Line(270,130)(278,122)
\Text(280,130)[l]{{\footnotesize leptons}}
\Text(195,100)[r]{$D$}\Text(305,100)[l]{$K$}
\Text(235,108)[b]{$c$}\Text(265,108)[b]{$s$}
\Text(250,65)[t]{(d)}
\end{picture}
\caption{Subprocesses from which the $V_{ud}$, $V_{us}$, $V_{cd}$, and
$V_{cs}$ elements of the CKM-Matrix are determined (see text).}
\label{fig:vcabibbo}
\end{center}
\end{figure}

\begin{itemize}
\item $\mathbf{V_{ud}}$:
\begin{equation}
|V_{ud}|=0.9736\pm 0.0010\ .
\label{eq:vud}\end{equation}
This is the best determined of the elements and is obtained by
studying superallowed $\beta$-decays in nuclei (see
fig.~\ref{fig:vcabibbo}(a)). Since the publication of the 1996
particle data book, a result with a smaller error has been presented,
$|V_{ud}|=0.9740\pm 0.0005$~\cite{chalkriver}.
\item $\mathbf{V_{us}}$:
\begin{equation}
|V_{us}|=\lambda = 0.2205\pm 0.0018\ .
\label{eq:vus}\end{equation}
This matrix element is obtained from semileptonic decays of $K$-mesons
($K^+\to\pi^0e^+\nu_e$, $K^0_L\to\pi^-e^+\nu_e$, see
fig.~\ref{fig:vcabibbo}(b)\,) or of hyperons. In the $K$-meson decays,
parity symmetry implies that only the vector component of the $V{-}A$
current contributes to the decays. Since the momentum transfer to the
$\pi$-meson is small, and the vector current for degenerate quarks
is conserved, the uncertainties due to strong interaction effects are
small and can be estimated using chiral perturbation theory.
\item $\mathbf{V_{cd}}$:
\begin{equation}
|V_{cd}|= 0.224\pm 0.016\ .
\label{eq:vcd}\end{equation}
The $V_{cd}$ element is obtained from charm production in deep
inelastic neutrino (or antineutrino) nucleon scattering, see
fig.~\ref{fig:vcabibbo}(c).
\item $\mathbf{V_{cs}}$:
\begin{equation}
|V_{cs}|= 1.01\pm 0.18\ .
\label{eq:vcs}\end{equation}
This matrix element can be determined from semileptonic decays of
charmed mesons (see fig.~\ref{fig:vcabibbo}(d)). In this case the
large difference in the masses of the quarks make it difficult to
estimate the strong interaction effects accurately (see
sec.~\ref{subsec:vcb}), and this is the reason for the relatively
large error.
\end{itemize}
The errors in eqs.~(\ref{eq:vud})-(\ref{eq:vcs}) are those in the
measurements of these matrix elements themselves. Comparing these
uncertainties with those in eq.~(\ref{eq:ckmvalues}), where the
constraints of unitarity have been imposed (assuming just three
generations), we see how much these constraints reduce the errors.

\subsection{Flavour Changing Neutral Currents}
\label{subsec:fcnc}
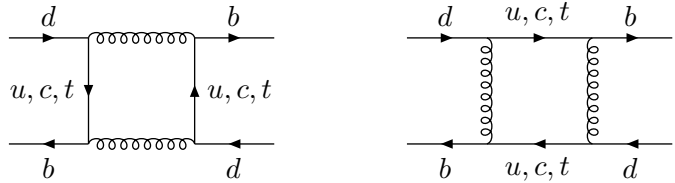
\begin{figure}[th]
\begin{center}
\begin{picture}(250,70)(0,45)
\ArrowLine(10,100)(40,100)\Gluon(40,100)(80,100){2}{8}
\ArrowLine(80,100)(110,100)
\ArrowLine(40,60)(10,60)\Gluon(40,60)(80,60){2}{8}
\ArrowLine(110,60)(80,60)
\ArrowLine(40,100)(40,60)\ArrowLine(80,60)(80,100)
\Text(25,105)[b]{$d$}\Text(95,105)[b]{$b$}
\Text(25,55)[t]{$b$}\Text(95,55)[t]{$d$}
\Text(35,80)[r]{$u,c,t$}\Text(85,80)[l]{$u,c,t$}
\ArrowLine(160,100)(190,100)\ArrowLine(190,100)(230,100)
\ArrowLine(230,100)(260,100)
\ArrowLine(190,60)(160,60)\ArrowLine(230,60)(190,60)
\ArrowLine(260,60)(230,60)
\Gluon(190,100)(190,60){2}{8}\Gluon(230,60)(230,100){2}{8}
\Text(175,105)[b]{$d$}\Text(245,105)[b]{$b$}
\Text(175,55)[t]{$b$}\Text(245,55)[t]{$d$}
\Text(210,105)[b]{$u,c,t$}\Text(210,55)[t]{$u,c,t$}
\end{picture}
\caption{{\emph Box} diagrams which contribute to the process of $B^0$-$\bar B^0$
mixing.}
\label{fig:box}
\end{center}
\end{figure}

\begin{figure}[th]
\begin{center}
\begin{picture}(430,130)(10,15)
\ArrowLine(10,100)(40,100)\ArrowLine(40,100)(70,100)\ArrowLine(70,100)(100,100)
\ArrowLine(100,100)(130,100)
\GlueArc(70,100)(30,180,0){3}{20}\Gluon(70,100)(70,60){2}{8}
\Text(70,105)[b]{$u,c,t$}\Text(20,95)[t]{$d$}\Text(120,95)[t]{$s$}
\Text(70,140)[b]{${\scriptstyle W}$}\Text(70,55)[t]{$Z^0,\gamma$}
\Text(70,30)[t]{(a)}
\ArrowLine(160,120)(190,120)\ArrowLine(190,120)(250,120)
\ArrowLine(250,120)(280,120)
\Gluon(190,120)(220,100){2}{7}\Gluon(250,120)(220,100){2}{7}
\Gluon(220,100)(220,60){2}{8}
\Text(220,125)[b]{$u,c,t$}\Text(170,115)[t]{$d$}\Text(270,115)[t]{$s$}
\Text(202,105)[r]{${\scriptstyle W}$}\Text(238,105)[l]{${\scriptstyle W}$}
\Text(220,55)[t]{$Z^0,\gamma$}\Text(220,30)[t]{(b)}
\ArrowLine(310,100)(340,100)\ArrowLine(340,100)(370,100)\ArrowLine(370,100)(400,100)
\ArrowLine(400,100)(430,100)
\GlueArc(370,100)(30,180,0){3}{20}\Photon(370,100)(370,60){2}{8}
\Text(370,105)[b]{$u,c,t$}\Text(320,95)[t]{$d$}\Text(420,95)[t]{$s$}
\Text(370,140)[b]{${\scriptstyle W}$}\Text(370,55)[t]{$\mathcal{G}$}
\Text(370,30)[t]{(c)}
\end{picture}
\caption{Examples of {\emph penguin} diagrams which contribute to the
FCNC process $d\to s$. Diagrams (a) and (b) are \emph{electroweak
penguins} graphs, whereas diagram (c) is a \emph{gluonic penguin}
graph.}
\label{fig:penguin}
\end{center}
\end{figure}
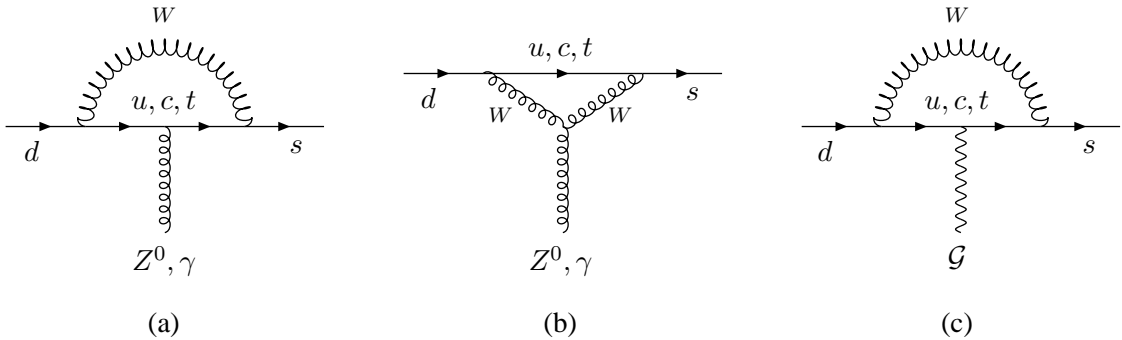

In the Standard Model, unitarity implies that there are no Flavour Changing
Neutral Current (FCNC) reactions at tree level, i.e. that there are
no vertices of the type:
\begin{center}
\begin{picture}(210,40)(0,-10)
\ArrowLine(0,20)(30,20)\ArrowLine(30,20)(60,20)
\ArrowLine(150,20)(180,20)\ArrowLine(180,20)(210,20)
\Gluon(30,20)(30,-10){2.5}{4}
\Gluon(180,20)(180,-10){2.5}{4}
\Text(15,25)[b]{$b$}\Text(45,25)[b]{$s$}
\Text(165,25)[b]{$u$}\Text(195,25)[b]{$c$}\Text(230,-10)[c]{.}
\end{picture}
\end{center}
Quantum loops can, however, generate FCNC reactions, through {\em box}
diagrams (see fig.~\ref{fig:box}) or {\em penguin} diagrams (see
fig.~\ref{fig:penguin}), and we will discuss some of the physical
processes induced by these loop-effects in the following lectures.

As a consequence of the ``GIM''-mechanism~\cite{gim} all the FCNC
vertices vanish in the limit of degenerate quark masses. This is a
consequence of the unitarity of the CKM-matrix. For example, consider
the vertex in the diagram of fig.~\ref{fig:penguin}(a), in the
hypothetical situation in which there is a ``horizontal'' symmetry
such that $m_u=m_c=m_t$. In this case the contribution from each of
the positively charged quarks in the loop is simply proportional to
the corresponding CKM-matrix elements, so that the total contribution
is
\begin{equation}
V_{ud}V_{us}^* + V_{cd}V_{cs}^* + V_{td}V_{ts}^*  
\label{eq:gim}\end{equation}
which vanishes by unitarity of the CKM-matrix. A similar argument
holds for all the FCNC vertices.

Of course, in reality the masses of the quarks are not equal and the
FCNC vertices are not zero. Depending on the process being studied,
however, the GIM-mechanism may lead to a suppression of the amplitude.

\subsection{$P$, $C$ and $CP$ Symmetries}
\label{subsec:cp}

Symmetries play a fundamental r\^ole in physics in general and in
particle physics in particular. In this subsection I briefly introduce
the violation of $CP$-symmetry, which will be a focus of many of the
studies at $B$-physics facilities in the next few years.

\paragraph{\bf Parity:} The parity transformation reverses the sign of spatial
coordinates, $(\vec x, t)\to (-\vec x, t)$. The vector and axial-vector fields
transform as:\begin{equation}
V_\mu(\vec x, t)\to V^\mu(-\vec x, t)\ \ \textrm{and}\ \ \ 
A_\mu(\vec x, t)\to -A^\mu(-\vec x, t)\ ,
\label{eq:pva}\end{equation}
and the vector and axial-vector currents transform similarly.
Left-handed components of fer\-mions $(\frac{1}{2}(1-\gamma^5)\psi)$
transform into right-handed ones $(\frac{1}{2}(1+\gamma^5)\psi)$, and
vice-versa. Since weak interactions only involve the left-handed
components, parity is not a good symmetry of the weak force, in
distinction to the strong and electromagnetic forces (QCD and QED are
invariant under parity transformations).

\paragraph{\bf Charge Conjugation:} For free fields ($\phi(x)$ say) we
can perform a fourier decomposition, with coefficients which contain
annihilation and creation operators:
\begin{equation}
\phi(x)=\ \cdots\  a\  +\  \cdots\  b^\dagger\ ,
\label{eq:cdef}\end{equation}
where $a$ and $b^\dagger$ represent the annihilation and creation
operators for the particle and antiparticle. The charge conjugation
transformation is the interchange of $a$ and $b$. In an interacting
field theory this is not in general the same as interchanging physical
particles and antiparticles for which we need the $CPT$ combined
transformation ($T$ is time reversal transformation). Nevertheless
$C$-tranformations are also interesting in field theories. Under $C$
the currents transform as follows:
\begin{equation}
\bar\psi_1\gamma_\mu\psi_2\to -\bar\psi_2\gamma_\mu\psi_1\ \ \
\textrm{and}\ \ \ \bar\psi_1\gamma_\mu\gamma_5\psi_2\to
\bar\psi_2\gamma_\mu \gamma_5\psi_1\ ,
\label{eq:cva}\end{equation}
where $\psi_i$ represents a spinor field of type (flavour or lepton
species) $i$.

\paragraph{$\mathbf{CP:}$} Under the combined $CP$-transformation, 
the currents transforms as:
\begin{equation}
\bar\psi_1\gamma_\mu\psi_2\to -\bar\psi_2\gamma^\mu\psi_1\ \ \
\textrm{and}\ \ \ \bar\psi_1\gamma_\mu\gamma_5\psi_2\to
-\bar\psi_2\gamma^\mu \gamma_5\psi_1\ ,
\label{eq:cpva}\end{equation}
where the fields on the left (right) hand side are evaluated at $(\vec x,t)$
(\,$(-\vec x,t)$\,). Consider now a charged current interaction:
\begin{equation}
(W_\mu^1-iW^2_\mu)\,\bar U^i\gamma^\mu(1-\gamma^5)V_{ij} D^j
+ (W_\mu^1+iW^2_\mu)\,\bar D^j\gamma^\mu(1-\gamma^5)V_{ij}^* U^i\ ,
\label{eq:ccint}\end{equation}
where $W$ represents the field of the $W$-boson and $U^i$ and $D^j$ are
up and down type quarks of flavours $i$ and $j$ respectively. Under a
$CP$ transformation, the interaction term in eq.~(\ref{eq:ccint})
transforms to:
\begin{equation}
(W_\mu^1 + i W^2_\mu)\,\bar D^j\gamma^\mu(1-\gamma^5)V_{ij} U^i
+ (W_\mu^1 - iW^2_\mu)\,\bar U^i\gamma^\mu(1-\gamma^5)V_{ij}^* D^j\ ,
\label{eq:cpcc}\end{equation}
where the parity transformation on the coordinates is implied. Comparing
eqs.~(\ref{eq:ccint}) and (\ref{eq:cpcc}), we see that $CP$-conservation
requires the CKM-matrix $V_{ij}$ to be real (or more strictly that any
phases must be absorbable into the definition of the quark fields). If
the only source of $CP$-violation in nature is the phase in the
CKM-matrix, then at least three generations of quarks are required (see
subsec.~\ref{subsec:ckm}). A central element of the research of the
forthcoming generation of experiments in $B$-physics will be to
determine whether the phase in the CKM-matrix is the only (or the main)
source of $CP$-violation.

$CP$-violation appears to be small compared to the strength of the weak
interaction, so that $CP$ is a fairly good symmetry. We recall that the
presence of CP-violation is one of the Sakharov conditions for the
creation of the baryon-antibaryon asymmetry of the universe.

\subsection{Leptonic Decays of Mesons}
\label{subsec:fp}

As already mentioned, the main difficulty in making predictions for weak
decays of hadrons is in controlling the non-perturbative
strong-interaction effects. We will encounter this problem several times
in subsequent lectures, but, as an introduction, I now discuss it
briefly in a particularly simple situation, that of the leptonic decays
of pseudoscalar mesons in general, and of the $B$-meson in particular.

\begin{figure}[ht]
\begin{center}
\begin{picture}(180,50)(0,15)
\SetWidth{2}\ArrowLine(10,41)(50,41)
\SetWidth{0.5}
\Oval(80,41)(15,30)(0)
\ArrowLine(79,56)(81,56)\ArrowLine(81,26)(79,26)
\GCirc(50,41){7}{0.8}
\Gluon(110,41)(150,41){3}{7}
\ArrowLine(150,41)(167,51)\ArrowLine(167,31)(150,41)
\Text(20,35)[tl]{\boldmath$B^-$}
\Text(80,62)[b]{$b$}\Text(80,20)[t]{$\bar u$}
\Text(172,53)[l]{$l^-$}\Text(172,30)[l]{$\bar\nu$}
\Text(132,48)[b]{$W$}
\end{picture}
\caption{Diagram representing the leptonic decay of the $B$-meson.}
\label{fig:fb}
\end{center}
\end{figure}
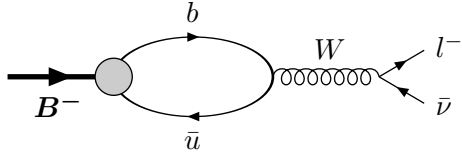

The leptonic decay of a pseudoscalar meson (e.g. the $B$-meson) is
illustrated in the diagram of Fig.~\ref{fig:fb}. All the QCD effects
corresponding to the strong interactions between the quarks in
Fig.~\ref{fig:fb} are contained in the matrix element of the $V{-}A$
hadronic current:
\begin{equation}
\langle 0|\,\bar b\gamma^\mu(1-\gamma^5)u\,|B(p)\rangle\ .
\label{eq:bldecay}\end{equation}
Lorentz and parity symmetries imply that the matrix element of the
vector component of the $V{-}A$ current vanishes ($\langle 0|\,\bar
b\gamma^\mu u\,|B(p)\rangle=0$). This component is an axial-vector,
and there is no axial-vector which can be constructed from the single
vector we have at our disposal ($p$) and the invariant tensors
($g^{\mu,\nu}$ and $\varepsilon^{\mu\nu\rho\sigma}$). These symmetries
also imply that the matrix element of the axial component of the
$V{-}A$ current is a Lorentz-vector, and hence that it is proportional
to $p^\mu$:
\begin{equation}
\langle 0|\,\bar b\gamma^\mu\gamma^5 u\,|B(p)\rangle=if_Bp^\mu\ .
\label{eq:fbdef}\end{equation}
The constant of proportionality, $f_B$, is called the \emph{decay
constant} of the $B$-meson. Thus all the QCD-effects for leptonic
decays of $B$-mesons are contained in a single number, $f_B$. An
identical discussion holds for other pseudoscalar mesons
($\pi,K,\cdots$). The normalisation used throughout these lectures,
corresponds to $f_\pi\simeq 132$~MeV ($f_B$ is unknown at
present). For leptonic decays of vector mesons, space-time symmetries
also imply that the strong interaction effects are contained in a
single decay constant. Thus, the r\^ole of quantitative
non-perturbative methods, such as lattice QCD (discussed in
sec.~\ref{subsec:lattice} below) or QCD sum-rules, is to provide the
framework for the evaluation of matrix elements such as that in
eq.~(\ref{eq:bldecay}).

\subsection{Operator Product Expansions and Effective Hamiltonians}
\label{subsec:ope}

Even though this is a course in flavour physics, we cannot avoid the
fact that the quarks are interacting strongly and hence that QCD
effects must be considered. Indeed, as already mentioned above, our
rather primitive control of non-perturbative QCD is the principal
source of uncertainty in interpeting experimental data of weak decays
and in determining the fundamental parameters of the Standard Model.

In Paolo Nason's lecture course~\cite{nason}, we have seen how the
property of \emph{asymptotic freedom}, which states that the
interactions of quarks and gluons become weaker at short distances,
enables us to use perturbation theory to make predictions for a wide
variety of short-distance (or light-cone) dominated processes. For
separations $|x|\ll \Lambda_{QCD}^{-1}$ (\,$|x|< 0.1$~fm say) or
corresponding momenta $|p|\gg \Lambda_{QCD}$ (\,$|p|>2$~GeV say), we
can use perturbation theory. The natural scale of strong interaction
physics is of $O(1~\textrm{fm})$, however, and so in general, and for
most of the processes discussed in this course, non-perturbative
techniques must be used.

\begin{figure}[th]
\begin{center}
\begin{picture}(350,100)(0,20)
\Line(30,100)(90,100)\Line(30,60)(90,60)
\Gluon(60,100)(60,60){2.5}{5}
\Text(25,100)[r]{$s$}\Text(25,60)[r]{$\bar d$}
\Text(95,100)[l]{$u$}\Text(95,60)[l]{$\bar u$}
\Text(70,80)[l]{$W$}\Text(60,40)[t]{(a)}
\Line(240,100)(340,100)\Line(240,60)(340,60)
\Gluon(310,100)(310,60){2.5}{5}
\Photon(270,100)(270,60){1.5}{8}
\Text(235,100)[r]{$s$}\Text(235,60)[r]{$\bar d$}
\Text(345,100)[l]{$u$}\Text(345,60)[l]{$\bar u$}
\Text(320,80)[l]{$W$}\Text(290,40)[t]{(b)}
\Text(262,80)[r]{${\cal G}$}
\end{picture}
\caption{Diagrams contributing to hadronic decays of $K$-mesons}
\label{fig:kdecay}
\end{center}
\end{figure}
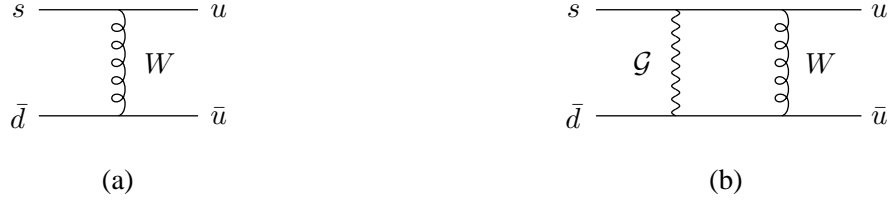

As an example, consider a decay of a $K$-meson into two pions, for which
a tree-level diagram of the quark subprocess is depicted in
Fig.~\ref{fig:kdecay}(a). The amplitude is proportional to
\begin{equation}
\frac{G_F}{\sqrt{2}}\,V_{ud}^*V_{us}\, \langle\pi\pi|(\bar
d\gamma^\mu(1-\gamma^5)u)\,(\bar u\gamma_\mu(1-\gamma^5) s)|K\rangle\ ,
\label{eq:me1}\end{equation}
i.e. it is proportional to the matrix element of the operator
\begin{equation}
O_1 = (\bar d\gamma^\mu(1-\gamma^5)u)\,(\bar u\gamma_\mu(1-\gamma^5) s)
\label{eq:o1def}\end{equation}
between $|K\rangle$ and $|\pi\pi\rangle$ states.
A one-loop (i.e. one-gluon exchange) correction to this diagram is shown
in Fig.~\ref{fig:kdecay}(b). This clearly generates a second operator,
$(\bar dT^a\gamma^\mu(1-\gamma^5)u)\,(\bar uT^a\gamma_\mu(1-\gamma^5)
s)$~\footnote{The $T^a$'s are the eight matrices, representing the
generators of the $SU(3)$ colour group in the fundamental
representation. The coupling of quarks and gluons is proportional to the
elements of these matrices.}, which, using Fierz identities can be
written as a linear combination of $O_1$ and $O_2$ where
\begin{equation}
O_2=(\bar d\gamma^\mu (1-\gamma^5)s)\,(\bar u \gamma_\mu(1-\gamma^5)u)\, .
\end{equation}
This discuusion can be generalized to higher orders, as I now explain. 

Using the operator product expansion (OPE) one writes the amplitude
for a weak decay process from an initial state $\mathcal{I}$ to a
final state $\mathcal{F}$ in the form:
\begin{equation}
T_{\mathcal FI} = \frac{G_F}{\sqrt{2}}\,V_{CKM}\,\sum_{i}C_i(\mu)
\langle \mathcal{F}|\,O_i(\mu)\,|\mathcal{I}\rangle\ . 
\label{eq:ope}\end{equation}
$\mu$ is the renormalization scale at which the composite operators
$O_i$ are defined and $V_{CKM}$ is the appropriate product of
CKM-matrix elements. The expansion (\ref{eq:ope}) is very
convenient. The non-perturbative QCD effects are contained in the
matrix elements of the operators $O_i$, which are independent of the
large momentum scale, in this case of $M_W$. The Wilson coefficient
functions $C_i(\mu)$ are independent of the states $\mathcal{I}$ and
$\mathcal{F}$ and can be calculated in perturbation theory. Since
physical amplitudes manifestly do not depend on $\mu$, the
$\mu$-dependence in the operators $O_i(\mu)$ is cancelled by that in
the coefficient functions $C_i(\mu)$. The \emph{effective Hamiltonian} 
for weak decays can then be written in the form:
\begin{equation}
\mathcal{H}_{\mathrm{eff}} \equiv
\frac{G_F}{\sqrt{2}}\,V_{CKM}\,\sum_{i}C_i(\mu)\,O_i(\mu)\ .
\end{equation}

In order to gain a little further intuition into the origin of
eq.(\ref{eq:ope}) consider the one-loop graph of
Fig.~\ref{fig:kdecay}(b). Dimensional counting at large loop momenta
$k$ readily shows that the graph is ultra-violet convergent (in the
Feynman gauge, say):
\begin{equation}
\int_{k \mathrm{\ large}}\frac{1}{k}\ \frac{1}{k}\ \frac{1}{k^2}\
\frac{1}{k^2-M_W^2}\ d^4k\ ,
\label{eq:powercounting}\end{equation}
where the first two factors in the integrand ($1/k$) correspond to the
two quark propagators, the third ($1/k^2$) to the gluon propagator and
the final factor to the propagator of the $W$-boson. It can be deduced
from eq.(\ref{eq:powercounting}) that the contribution from this graph
contains a term proportional to $\log (M_W^2/p^2)$, where p is some
low-momentum (infra-red) scale.

\begin{figure}[th]
\begin{center}
\begin{picture}(150,60)(0,50)
\Line(20,100)(90,80)\Line(90,80)(130,100)
\Line(20,60)(90,80)\Line(90,80)(130,60)
\GCirc(90,80){10}{0.7}
\Photon(45,92.86)(45,67.14){1.5}{7}
\Text(15,100)[r]{$s$}\Text(15,60)[r]{$\bar d$}
\Text(135,100)[l]{$u$}\Text(135,60)[l]{$\bar u$}
\Text(38,80)[r]{$\mathcal{G}$}
\Text(92,65)[t]{$O_1$}
\end{picture}
\caption{One-loop correction to the Green function of the operator $O_1$.}
\label{fig:operator}
\end{center}
\end{figure}
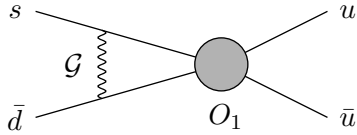

Consider now the one-loop graph of Fig.~\ref{fig:operator}, depicting
a correction to a matrix element of the operator $O_1$ defined in
eq.~(\ref{eq:o1def}). Now the $W$-propagator is absent, and the power
counting in the large momentum region gives:
\begin{equation}
\int_{k \mathrm{\ large}}\frac{1}{k}\ \frac{1}{k}\ \frac{1}{k^2}\
d^4k\ ,
\label{eq:powercounting2}\end{equation}
i.e. a logarithmically divergent contribution. After renormalization
at a scale $\mu$ there will be a contribution to this graph
proportional to $\log (\mu^2/p^2)$. In the low momentum
(non-perturbative) region (i.e. for $|k_\mu|\ll M_W$) the
contributions from the graphs in Figs.~\ref{fig:kdecay}(b) and
\ref{fig:operator} are the same. Thus we can split the factor of $\log
(M_W^2/p^2)$ from the graph of Fig.~\ref{fig:kdecay}(b) into two
parts; the first, $\log (\mu^2/p^2)$ is contained in the matrix
element of $O_1$, and the second, $\log (M_W^2/\mu^2)$, into the
coeficient function $C_1$. Since the infra-red behaviour of the graphs
in Figs.~\ref{fig:kdecay}(b) and \ref{fig:operator} are the same, the
coefficient functions are independent of the treatment of the
low-momentum region. Moreover, in order to calculate the coefficient
functions, we can choose any convenient external states, and in
practice one chooses quark or gluon states. We refer to the evaluation
of the coefficient functions as the process of \emph{matching}.

In $K$-decays it is natural, although not necessary, to choose $\mu$
to be as small as one can without invalidating perturbation theory. In
this way one might hope to avoid \emph{large logarithms}
($\log(\mu^2/p^2$)\,) which would spoil any insights we might have about
the operator matrix elements from non-relativistic quark models or
other approaches. Of course this implies the presence of large
logarithms of the type $\alpha_s^n\log^n(M_W^2/\mu^2)$ in the
coefficient functions, but these can be summed using the
renormalization group, leading to factors of the type:
\begin{equation}
\left[\frac{\alpha_s(M_W)}{\alpha_s(\mu)} \right]^{\gamma_0/2\beta_0}\ ,
\label{eq:leadinglogs}\end{equation}
where $\gamma_0$ is the one-loop contribution to the \emph{anomalous
dimension} of the operator (proportional to the coefficient of
$\log(\mu^2/p^2)$ in the evaluation of the graph of
fig.~\ref{fig:kdecay}(b)\,) and $\beta_0$ is the first term in the
$\beta$-function, ($\beta\equiv\partial
g/\,\partial\ln(\mu)\,=\,-\beta_0\,g^3/16\pi^2$)~\footnote{In general
when there is more than one operator contributing to the right hand
side of eq.~(\ref{eq:ope}), the expression in
eq.~(\ref{eq:leadinglogs}) is generalized into a matrix equation,
representing the mixing of the operators}.  Eq.~(\ref{eq:leadinglogs})
represents the sum of the \emph{leading logarithms}, i.e. the sum of
the terms $\alpha_s^n\log^n(M_W^2/\mu^2)$. For almost all the
important processes, the first (or even higher) corrections have also
been evaluated.

We end this subsection by making some further points about the use
of the OPE in weak decays and other hard processes:
\begin{itemize}
\item We shall see below that for some important physical quantities
(e.g. $\epsilon^\prime/ \epsilon$), there may be as many as ten
operators, whose matrix elements have to be estimated.
\item One may try to evaluate the matrix elements using some
non-perturbative computational technique, such as lattice simulations or
QCD sum-rules (see below), or to determine them from experimental data.
In the latter case, if there are more measurements possible than unknown
parameters (i.e. matrix elements), then one is able to make predictions.
\item In weak decays the large scale, $M_W$, is of course fixed. For
other processes, most notably for deep-inelastic lepton-hadron
scattering, the OPE is useful in computing the behaviour of the
amplitudes with the large scale (e.g. with the momentum transfer).
\end{itemize}

\subsection{The Heavy Quark Effective Theory - HQET}
\label{subsec:hqet}

Most of the important physical properties of the hydrogen atom are
independent of the mass of the nucleus. Analogous features hold in the
physics of heavy hadrons, where by \emph{heavy} I mean that
$m_Q\gg\Lambda_{QCD}$ and $m_Q$ is the mass of the heavy quark. During
the last few years the construction and use of the \emph{Heavy Quark
Effective Theory} (HQET), has proved invaluable in the study of heavy
quark physics. In this subsection I briefly introduce the main features
of the HQET, which will be used in the following lectures.

Consider the propagator of a (free) heavy quark: \begin{center}
\begin{picture}(0,10)(100,50)
\ArrowLine(10,50)(70,50)\Text(40,45)[t]{$p$}
\Text(80,50)[l]{$=\ i\,\frac{\not{\hspace{0.5pt}p}+m}{p^2-m_Q^2+i\epsilon}$
.}\end{picture} \end{center}
If the momentum of the quark $p$ is not far from its mass shell,
\begin{equation}
p_\mu = m_Q v_\mu + k_\mu\ ,
\end{equation}
where $|k_\mu|\ll m_Q$ and $v_\mu$ is the (relativistic)
four velocity of the hadron containing the heavy quark ($v^2=1$), then
\begin{center}\begin{picture}(0,10)(100,50)
\ArrowLine(10,50)(70,50)\Text(40,45)[t]{$p$}
\Text(80,50)[l]{$=\ i\,\frac{1+\not{\hspace{0.5pt}v}}{2}
\frac{1}{v\cdot k + i\epsilon} + O\left(\frac{|k_\mu|}{m_Q}\right)$.}
\end{picture}\end{center}
$(1+\not{\hspace{-3.3pt}v})/2$ is a projection operator, projecting out
the \emph{large} components of the spinors. This propagator can be
obtained from the gauge-invariant action
\begin{equation}
\mathcal{L}_{HQET} = \bar h (iv\cdot D)
\frac{1+\not{\hspace{-2pt}v}}{2} h\ ,
\label{eq:lhqet}\end{equation}
where $h$ is the spinor field of the heavy quark~\footnote{Of course
there are formal derivations of the action in eq.~(\ref{eq:lhqet}).}.
$\mathcal{L}_{HQET}$ is independent of $m_Q$, which implies the
existence of symmetries relating physical quantities corresponding to
different heavy quarks (in practice the $b$ and $c$ quarks). The light
degrees of freedom are also not sensitive to the spin of the heavy quark,
which leads to a spin-symmetry relating physical properties of heavy
hadrons of different spins. Consider, for example, the correlation
function:
\begin{equation}
\int d^3 x\,\langle 0 | J_H(x)\,J_H^\dagger(0)\,|0\rangle\ ,
\label{eq:cf}\end{equation}
where $J_H^\dagger$ ($J_H$) is an interpolating operator which can
create a heavy hadron $H$, which we take to be a pseudoscalar or
vector meson. The hadron is produced at rest, with four velocity
$v=(\vec 0, 1)$. We can use the interpolating operator $J_H = \bar
h\gamma^5 q$ for the pseudoscalar meson and $J_H = \bar h\gamma^i q$
($i= 1,2,3$) for the vector meson. This means that the correlation
function will be identical in the two cases except for the factor
\begin{equation}
\gamma^5 \frac{1 + \gamma^0}{2} \gamma^5 = \frac{1 - \gamma^0}{2}
\label{eq:cfp}\end{equation}
when $H$ is a pseudoscalar meson, and
\begin{equation}
\gamma^i \frac{1 + \gamma^0}{2} \gamma^i = -3\,\frac{1 - \gamma^0}{2}
\label{eq:cfv}\end{equation}
when it is a vector meson. Since the correlation functions behave with
time as $\exp (-i M_H t)$ , this implies that the pseudoscalar and
vector mesons are degenerate, up to relative corrections of
$O(\Lambda_{QCD}^2/m_Q)$:
\begin{equation}
M_P = M_V + O(\Lambda_{QCD}^2/m_Q)\ ,
\label{eq:degenerate} \end{equation}
where $M_P$ and $M_V$ are the masses of the pseudoscalar and vector
mesons respectively, or equivalently
\begin{equation}
M_V^2 - M_P^2 = \textrm{constant}\ .
\label{eq:degenerate2}\end{equation}
This relation is reasonably well satisfied experimentally:
\begin{equation}
M_{B^*}^2 - M_B^2 = 0.485\,\textrm{GeV}^2\ \ \
\textrm{and}\ \ \ M_{D^*}^2 - M_D^2 = 0.546\,\textrm{GeV}^2\ ,
\label{eq:degenerate3}\end{equation}
which is encouraging. However, I also mention in passing that the
light mesons, to which the HQET certainly does not apply, also
satisfy these relations numerically:
\begin{equation}
M_{K^*}^2 - M_K^2 = 0.552\,\textrm{GeV}^2\ \ \
\textrm{and}\ \ \ M_{\rho}^2 - M_\pi^2 = 0.571\,\textrm{GeV}^2\ .
\label{eq:degenerate4}\end{equation}

The usefulness of the HQET does not lie in the fact that the
\emph{residual} momenta $k$ are always much smaller than
$\Lambda_{QCD}$; indeed this is not true, in QCD one does have hard
gluons with arbitrarily large momenta. However the effects of hard
gluons can be evaluated in perturbation theory. The non-perturbative
strong interaction effects are the same in QCD and in the HQET (up to
corrections of $O(\Lambda_{QCD}/m_Q)$, which for the moment we
neglect), and so the HQET relations, such as that between the decay
constants of pseudoscalar and vector mesons, which can be deduced from
the above proportionality of correlation functions, are violated only
by perturbatively calculable corrections.

The principal application of the HQET is in $b$-physics, and to a lesser
extent to charm physics. In both cases the corrections of
$O(\Lambda_{QCD}/m_Q)$ are significant, and hence one would like to
compute these corrections. In practice, this gets progressively more
difficult, and it is a much debated point as to whether even the first
corrections (i.e. those of $O(\Lambda_{QCD}/m_Q)$ relative to the
leading terms) have been reliably calculated for any interesting
quantity. 

\section{Lecture 2: $V_{cb}$ and $V_{ub}$}

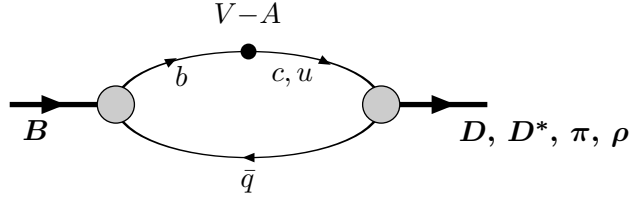
\begin{figure}
\begin{center}
\begin{picture}(180,70)(20,10)
\SetWidth{2}\ArrowLine(10,41)(43,41)
\Text(15,35)[tl]{\boldmath$B$}
\SetWidth{0.5}
\Oval(100,41)(20,50)(0)
\SetWidth{2}\ArrowLine(157,41)(190,41)
\Text(180,35)[tl]{\boldmath$D,\,D^*,\,\pi,\,\rho$}
\SetWidth{0.5}
\Vertex(100,61){3}
\GCirc(50,41){7}{0.8}\GCirc(150,41){7}{0.8}
\Text(75,48)[b]{$b$}\Text(117,48)[b]{$c,u$}
\Text(100,16)[t]{$\bar q$}
\Text(100,71)[b]{$V{-}A$}
\ArrowLine(101,21)(99,21)
\ArrowLine(70,57)(72,58)\ArrowLine(128,57.5)(130,56.5)
\end{picture}
\caption{Diagram representing the semileptonic decay of the $B$-meson.
$\bar q$ represents the light valence antiquark, and the black circle
represents the insertion of the $V$--$A$ current with the appropriate
flavour quantum numbers.}
\label{fig:sl}
\end{center}
\end{figure}

In this lecture I will discuss exclusive semileptonic decays of
$B$-mesons, in which the $b$-quark decays into a $c$- or $u$-quark, and
from which one can determine the $V_{cb}$ and $V_{ub}$ elements of the
CKM-matrix.
The decays are represented in Fig.~\ref{fig:sl}. It is
convenient to use space-time symmetries to express the matrix elements
in terms of invariant form factors (using the helicity basis for these
as defined below).  When the final state is a pseudoscalar meson $P$,
parity implies that only the vector component of the $V{-}A$ weak
current contributes to the decay, and there are two independent form
factors, $f^+$ and $f^0$, defined by\begin{equation}
\langle P(k)| V^\mu|B(p)\rangle  = 
  f^+(q^2)\left[(p+k)^\mu -
  \frac{m_B^2 - m_P^2}{q^2}\,q^\mu\right]  
+ f^0(q^2)\,\frac{m_B^2 - m_P^2}{q^2}\,q^\mu\ ,
\label{eq:ffpdef}\end{equation}
where $q$ is the momentum transfer, $q=p{-}k$.  When the final-state
hadron is a vector meson $V$, there are four independent form factors,
$V$, $A_1$, $A_2$ and $A$:
\begin{eqnarray}
\langle V(k,\varepsilon)| V^\mu|B(p)\rangle & = &
  \frac{2V(q^2)}{m_B+m_V}\epsilon^{\mu\gamma\delta\beta}
  \varepsilon^*_\beta p_\gamma k_\delta \label{eq:ffvvdef}\\ 
\langle V(k,\varepsilon)| A^\mu|B(p)\rangle  & = &  
  i (m_B {+} m_V) A_1(q^2) \varepsilon^{*\,\mu}
 - 
  i\frac{A_2(q^2)}{m_B{+}m_V} \varepsilon^*\!\dotprod p\,
  (p {+} k)^\mu \nonumber \\
& & \mbox{} + i \frac{A(q^2)}{q^2} 2 m_V 
  \varepsilon^*\!\dotprod p\, q^\mu\ , \label{eq:ffvadef}
\end{eqnarray}
where $\varepsilon$ is the polarization vector of the final-state
meson, and $q = p{-}k$.  Below we shall also discuss the form factor
$A_0$, which is given in terms of those defined above by $A_0 = A +
A_3$, with
\begin{equation}
A_3 = \frac{m_B + m_V}{2 m_V}A_1 - 
\frac{m_B - m_V}{2 m_V}A_2\ .
\label{eq:a3def}
\end{equation} 

\subsection{Semileptonic $B\to D$ and $B\to D^*$ Decays}
\label{subsec:vcb}

Semileptonic $B\to D^*$ and, more recently, $B \to D$ decays are used
to determine the $V_{cb}$ element of the CKM matrix. Heavy quark
symmetry is rather powerful in controlling the theoretical description
of these heavy-to-heavy quark transitions (see e.g. the review article
by Neubert for details and references~\cite{neubert}).

In the heavy quark limit, all six form factors in
eqs.~(\ref{eq:ffpdef})\,-\,(\ref{eq:ffvadef}) are related and there is just
one universal form factor $\xi(\omega)$, known as the Isgur--Wise (IW)
function, which contains all the non-perturbative QCD effects.
Specifically:
\begin{equation}
f^+(q^2) =  V(q^2) = A_0(q^2) = A_2(q^2)  =  \left[1 - \frac{q^2}{(m_B +
m_D)^2}\right]^{-1} A_1(q^2) = \frac{m_B+m_D}
{2\sqrt{m_Bm_D}}\,\xi(\omega)\ ,
\label{eq:iw}\end{equation}
where $\omega = v_B\dotprod v_D$ is the velocity transfer variable
($v_{B,D}$ are the four velocities of the corresponding mesons). Here
the label $D$ represents the $D$- or $D^*$-meson as appropriate
(pseudoscalar and vector mesons are degenerate in this leading
approximation). Vector current conservation implies that the IW-function
is normalized at zero recoil, i.e.\ that $\xi(1) =1$. This property is
particularly important in the extraction of $V_{cb}$.

To get some insights into the origin of the heavy quark relations above,
let us consider $B\to D$ decays. We can rewrite eq.~(\ref{eq:ffpdef}) in
terms of the form factors $f^+$ and $f^-$ (instead of using the helicity
basis as in eq.~(\ref{eq:ffpdef})\,):
\begin{equation} \langle D(p^\prime)| \bar c\gamma^\mu
b|B(p)\rangle  = f^+(q^2)(p+p^\prime)^\mu - f^-(q^2)(p-p^\prime)^\mu\ , 
\label{eq:ffpdef2}\end{equation} where $q=p-p^\prime$.
Defining the four-velocities $v$ and $v^\prime$ by $p=m_B\,v$ and
$p^\prime=m_D\,v^\prime$, eq.~(\ref{eq:ffpdef2}) can be rewritten as:
\begin{eqnarray}
\langle D(v^\prime)|\,\bar c\gamma^\mu b\,|B(v)\rangle & = & \frac{1}{2}
\left[ (m_B+m_D)f^+(q^2) - (m_B-m_D)f^-(q^2)\right](v+v^\prime)^\mu 
\nonumber\\ 
& & \hspace{0.5in}+\frac{1}{2}
\left[ (m_B-m_D)f^+(q^2) - (m_B+m_D)f^-(q^2)\right](v-v^\prime)^\mu \ .
\label{eq:ffpdef3}\end{eqnarray}
Now in the HQET:
\begin{equation}
\sqrt{m_Bm_D}\,\langle P(v^\prime)|\,\bar h_{v^\prime}\gamma^\mu h_v
\,|P(v)\rangle = \sqrt{m_Bm_D}\,\xi(\omega)\, (v + v^\prime)^\mu \ ,
\label{eq:ffphqet}\end{equation}
where $\omega=v\dotprod v^\prime$, and the projection operators (e.g.
$(1+\hspace{-7pt}\not\!\!\!v)/2$) 
have been absorbed into the heavy quark fields. $P$
represents a heavy pseudoscalar meson, and the factors $\sqrt{m_Bm_D}$
are chosen in order to introduce a mass-independent normalization of
states (so that the matrix elements are independent of any masses).
There is no term proportional to $(v-v^\prime)^\mu$ on the right hand
side of eq.~(\ref{eq:ffphqet}), since $\bar h_{v^\prime} (\not\! v -
\not\! v^\prime)h_v=0$, and so we obtain the relations:
\begin{equation}
f^-=\frac{m_B-m_D}{m_B+m_D}f^+\ \ \ \textrm{and}\ \ \
f^\pm = \frac{m_B\pm m_D}{2\sqrt{m_Bm_D}}\,\xi(\omega)\ .
\end{equation}
To understand the normalization condition $\xi(1)=1$, consider the
forward matrix element:
\begin{equation}
\langle B(p)|\,\bar b\gamma^\mu b\,|B(p)\rangle = 2p^\mu f^+(0) =
2 p^\mu\ ,
\label{eq:xi1}\end{equation}
where the last relation follows from the conservation of the current
$\bar b\gamma^\mu b$ which implies that $f^+(0)=1$. In the HQET this
matrix element is equal to $m_B\,2v^\mu\,\xi(1)$, and so $\xi(1)=1$. The
spin symmetry relates the form-factors of $B\to D^*$ decays to
$\xi(\omega)$.

The relations in eq.~(\ref{eq:iw}) are valid up to perturbative and
power corrections. The precision with which $V_{cb}$ can be extracted
is limited by the theoretical uncertainties in the evaluation of these
corrections. Nevertheless we are in the fortunate situation that it is
uncertainties in {\it corrections} (which are therefore small) which
control the precision.

Allowing for these corrections to the results in the heavy quark limit,
one writes the decay distribution for $B\to D^*$ decays as:
\begin{equation}
\frac{d\Gamma}{d\omega} =  
 \frac{G_F^2}{48\pi^3} (m_B{-}m_{D^*})^2 m_{D^*}^3
    \sqrt{\omega^2{-}1}\,(\omega{+}1)^2
\left[ 1 + \frac{4\omega}{\omega{+}1} 
     \frac{m_B^2-2\omega m_Bm_{D^*}+m_{D^*}^2}{(m_B-m_{D^*})^2}
    \right]
    |V_{cb}|^2\, {\cal F}^2(\omega)\ ,
\label{eq:distr}\end{equation}
where ${\cal F}(\omega)$ is the IW-function combined with perturbative
and power corrections. It is convenient theoretically to
consider this distribution near $\omega = 1$, in order to exploit the
normalization condition $\xi(1) = 1$. In this case there are no $O(1/m_Q)$
corrections (where $Q= b$ or $c$) by virtue of Luke's theorem~\cite{luke},
so that the expansion of $\mathcal{F}(1)$ begins like:
\begin{equation}
{\cal F}(1) = \eta_A\left( 1 + 0\,\frac{\Lambda_{QCD}}{m_Q} + 
c_2\frac{\Lambda^2_{QCD}}{m_Q^2} + \cdots\right)\, ,
\label{eq:fexpansion}\end{equation}
where $\eta_A$ represents the perturbative corrections.
The one-loop contribution to $\eta_A$ has been known for some time now,
whereas the two-loop contribution was evaluated last year, with the
result~\cite{czarnecki}:
\begin{equation}
 \eta_A = 0.960\pm 0.007\ ,
\end{equation}
where I have taken the value of the two loop contribution as an
estimate of the error.

The power corrections are much more difficult to estimate reliably.
Neubert has recently combined the results of
refs.~\cite{fn,mannel,suv} to estimate that the $O(1/m_Q^2)$ terms in
the parentheses in eq.~(\ref{eq:fexpansion}) are about $-0.055\pm
0.025$ and that ${\cal F}(1) = 0.91 (3)\,$.  Bigi, Shifman and
Uraltsev~\cite{bsu}, consider the uncertainties to be bigger and
obtain $0.91(6)$. Combining the latter, more cautious, theoretical
value of ${\cal F}(1)$, with the experimental result~\cite{diciaccio}
${\cal F}(1)|V_{cb}|= (34.3\pm 1.6)10^{-3}$, readily gives
$|V_{cb}|=(37.7\pm 1.8_{exp}\pm 2.5_{th})10^{-3}$.

Here I am only discussing exclusive decays. Buras and Fleischer, perform
an analysis of both exclusive and inclusive semileptonic decays and
quote $V_{cb}=(40\pm3)\,10^{-3}$ as their best value~\cite{bf}.

$|V_{cb}|$ gives the scale of the unitarity triangle:
\begin{equation}
\lambda\,|V_{cb}| = \lambda^3 A\ .
\label{eq:utvcb}\end{equation}

\subsection{Semileptonic $B\to \pi$ and $B\to\rho$ Decays}
\label{subsec:vub}

In this subsection we consider the heavy-to-light semileptonic decays
$B\to\rho$ and $B\to\pi$ which are now being used experimentally to
determine the $V_{ub}$ matrix element~\cite{lkg:ichep96,jrp:ichep96}.
Heavy quark symmetry is less predictive for heavy-to-light decays than
for heavy-to-heavy ones.  In particular, there is no normalization
condition at zero recoil corresponding to the condition $\xi(1)=1$,
which is so useful in the extraction of $V_{cb}$ (see
subsection~\ref{subsec:vcb}). The lack of such a condition puts a premium
on the results from nonperturbative calculational techniques, such as
lattice QCD or light-cone sum rules. Heavy quark symmetry does, however,
give useful scaling laws for the behaviour of the form factors with the
mass of the heavy quark ($m_Q$) at fixed $\omega$:
\begin{equation}
f^+, A_0,A_2,V\sim\sqrt{m_Q}\ ;\ A_1,f^0\sim\frac{1}{\sqrt{m_Q}}\ ;\ 
A_3\sim m_Q^{\frac{3}{2}}\ .
\label{eq:hqscaling}\end{equation}
These scaling relations are particularly useful in lattice simulations,
where the masses of the quarks are varied.  Moreover, the heavy quark
spin symmetry relates the $B\to V$ matrix elements~\cite{iw:hqet,gmm}
(where $V$ is a light vector particle) of the weak current and magnetic
moment operators, thereby relating the amplitudes for the two physical
processes $\btorho$ and $\btokstargamma$, up to $SU(3)$ flavour symmetry
breaking effects.  These relations also provide important checks on
theoretical, and in particular on lattice, calculations.

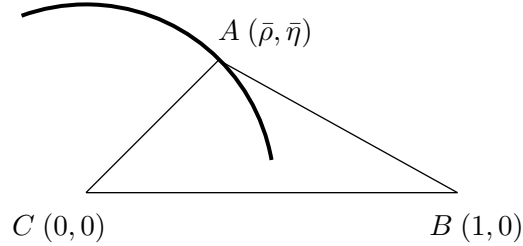
\begin{figure}[ht]
\begin{center}
\begin{picture}(160,90)(0,0)
\Line(10,10)(150,10)\Line(10,10)(60,60)\Line(60,60)(150,10)
\Text(60,65)[lb]{$A\ (\bar\rho,\bar\eta)$}
\Text(140,2)[lt]{$B\ (1,0)$}
\Text(0,2)[t]{$C\ (0,0)$}
\SetWidth{1.5}
\CArc(10,10)(71,10,110)
\end{picture}
\end{center}
\caption{A precise determination of $|V_{ub}/V_{cb}|$ would fix
the vertex $A$ to lie on a given circle centered on $C$ (schematically
represented by the solid line).}
\label{fig:vub2}\end{figure}

Recent compilations of results for $|V_{ub}/V_{cb}|$, include 0.06-0.11
obtained from exclusive decays only (or $0.08\pm 0.01_{\mbox{exp}} \pm
0.02_{\mbox{th}}$ from inclusive decays)~\cite{neubert,inclusive} 
and $0.08\pm 0.02$
from both inclusive and exclusive decays~\cite{bf}. In terms of the
parameters of the unitarity triangle (see fig.~\ref{fig:ut}):
\begin{equation}
CA = \frac{|V_{ud}V_{ub}^*|}{|V_{cd}V_{cb}^*|} =
\sqrt{\bar\rho^2 + \bar\eta^2} = (1-\frac{\lambda^2}{2})\frac{1}{\lambda}
\frac{|V_{ub}|}{|V_{cb}|}\simeq 4.44\,\frac{|V_{ub}|}{|V_{cb}|}\ .
\end{equation}
Thus a measurement of $|V_{ub}/V_{cb}|$ implies that the locus of
possible positions of the vertex $A$ is a circle centred on $C$ (see
fig.~\ref{fig:vub2}). In practice, of course, because of the theoretical
and experimental uncertainties, the circle is replaced by a band.

\subsection{Lattice QCD}
\label{subsec:lattice}

We have seen that the main difficulty in making predictions for weak
hadronic decays is our inability to control the long-distance strong
interaction effects present in these decays. The principal technique for
the evaluation of these non-perturbative effects is lattice QCD, and I
now digress from the main theme of this course to outline briefly  the
principles underlying lattice calculations of non-perturbative QCD
effects in weak decays and to discuss the main sources of uncertainty
present in these computations.

In subsection~\ref{subsec:ope} we have seen that, by
using the Operator Product Expansion, it is generally possible to
separate the short- and long-distance contributions to weak decay
amplitudes into Wilson coefficient functions and operator matrix
elements respectively. Thus, in order to evaluate the non-perturbative
QCD effects, it is necessary to compute the matrix elements of local
composite operators. This is achieved in lattice simulations, by the
direct computation of correlation functions of multi-local operators
composed of quark and gluon fields (in Euclidean space):
\begin{equation}
\langle0|\,O(x_1, x_2,\ldots,x_n)|0\rangle =\, \frac{1}{Z}
\int[DA_\mu][D\psi][D\bar\psi]\,e^{-S}\,O(x_1, x_2,\ldots,x_n)\ ,
\label{eq:vev}
\end{equation}
where $Z$ is the partition function 
\begin{equation}
Z=\int[DA_\mu][D\psi][D\bar\psi]\,e^{-S}\ ,
\label{eq:zdef}
\end{equation}
$S$ is the action and the integrals are over quark and gluon fields
at each space-time point. In eq.~(\ref{eq:vev}) $O(x_1, x_2,\ldots,x_n)$
is a multi-local operator; the choice of $O$ governs the physics which 
can be studied. 

We now consider the two most frequently encountered cases, for which
$n{=}2$ or $3$. As a first example, let $O(x_1, x_2)$ be the bilocal
operator
\begin{equation} O_2(x_1,x_2)=T\{J_H(x_1)J^\dagger_H(x_2)\}\ ,
\label{eq:o2}\end{equation}
where $J_H$ is an interpolating operator for the hadron $H$ whose
properties we wish to study. In lattice computations we evaluate the two
point correlation function
\begin{equation}
C_2(t_x)\equiv\sum_{\vec x} e^{i\vec p\dotprod \vec x}
\langle\,0\,|O_2(x, 0)|\,0\,\rangle\ .
\label{eq:c2def}\end{equation}
Inserting a complete set of states $\{|n\rangle\}$ we have:
\begin{eqnarray}
C_2(t_x) & = & \sum_n\sum_{\vec x} e^{i\vec p\dotprod \vec x}
\langle\,0\,|J_H(\vec x,t)|\,n\rangle\,
\langle n\,|\,J^\dagger_H(\vec 0,0)\,|\,0\,\rangle\ ,
\label{eq:c2exp0}\\ 
& = & \sum_{\vec x} e^{i\vec p\dotprod \vec x}
\langle\,0\,|J_H(\vec x,t)|\,H\rangle\,
\langle H\,|\,J^\dagger_H(\vec 0,0)\,|\,0\,\rangle\ +\cdots\ .
\label{eq:c2exp}\end{eqnarray}
In eq.~(\ref{eq:c2exp}), the ellipsis represents the contributions
from heavier states than $H$, which we assume to be the lightest
hadron which can be created by the operator $J_H^\dagger$. Using the
translation operator to move the argument of $J_H$ to zero, we
find~\footnote{The phase-space factor of 1/$2E_H$ in
eq.~(\ref{eq:c2exp2}) is implicitly included as part of the definition
of the summation over states in eqs.~(\ref{eq:c2exp0}) and 
(\ref{eq:c2exp}).}:
\begin{equation}
C_2(t_x)=\frac{e^{-E_Ht_x}}{2E_H}\,|\langle\,0\,|J_H(0)|\,h\,\rangle
|^2\ +\ \cdots\ ,
\label{eq:c2exp2}\end{equation}
where $E_H=\sqrt{M_H^2+\vec p^2}$. At large positive times $t_x$
the contribution from each heavier hadron, $H'$ with
mass $m_{H'}$ say, is suppressed by an exponential factor,
$\exp\big(-(E_{H'}-E_H)t_x\big)$, so that the contribution from the
lightest state $H$ is isolated. This is illustrated in the following
diagram:
\begin{center}
\begin{picture}(120,20)(0,10)
\ArrowLine(20,20)(100,20)\Text(60,25)[b]{$H$}
\GCirc(20,20){1}{0}\GCirc(100,20){1}{0}
\Text(20,14)[t]{$0$}\Text(100,14)[t]{$t_x$}\Text(120,14)[t]{.}
\end{picture}
\end{center}

In lattice simulations the correlation function $C_2$ is computed
numerically, by discretizing space-time (hence the word \emph{lattice}),
evaluating the functional integral in eq.~(\ref{eq:vev}) by Monte-Carlo
integration. By fitting the results to the expression in
eq.~(\ref{eq:c2exp2}) both the mass $M_H$ and the matrix element
$\langle\,0\,|J_H(0)|\,h\,\rangle$ can be
determined~\footnote{Frequently it is most convenient to evaluate the
correlation function with the hadron at rest, i.e. with $\vec p=0$.}. 

As an example consider the case in which $H$ is the $B$-meson and
$J_H$ is the axial current $A_\mu$ (with the flavour quantum numbers
of the $B$-meson). In this case one obtains the value of the leptonic
decay constant $f_B$, defined in eq.~(\ref{eq:fbdef}).

It will also be useful to consider three-point correlation functions:
\begin{equation}
C_3(t_x, t_y) = \sum_{\vec x,\vec y}
e^{i\vec p\dotprod\vec x} e^{i\vec q\dotprod\vec y}
\langle 0\,|\,J_2(\vec x, t_x)\, O(\vec y,t_y)\, J^\dagger_1(\vec 0, 0)\,
|\, 0\rangle\ ,
\label{eq:c3def}\end{equation}
where, $J_1$ and $J_2$ are the interpolating operators for hadrons
$H_1$ and $H_2$ respectively, $O$ is a local operator, and we have assumed
that $t_x>t_y>0$. Inserting complete sets of states between
the operators in eq.~(\ref{eq:c3def}) we obtain
\begin{eqnarray}
\lefteqn{C_3(t_x, t_y) = 
\frac{e^{-E_1t_y}}{2 E_1}\ \frac{e^{-E_2(t_x - t_y)}}{2 E_2}\,
\langle 0|J_2(\vec 0, 0)|H_2(\vec p, E_2)\rangle\times}\nonumber\\ 
& & \langle H_2(\vec p, E_2)|O(\vec 0,0)|H_1(\vec p {+} \vec q, E_1)\rangle\,
\langle H_1(\vec p {+} \vec q, E_1)|J^\dagger_1(\vec 0,0)|0\rangle+\cdots,
\label{eq:c3states}\end{eqnarray}
where $E_1=\sqrt{M_{H_1}^2 + (\vec p{+}\vec q)^2}$,
$E_2=\sqrt{M_{H_2}^2+\vec p^2}$ and the ellipsis represents the
contributions from heavier states.  The exponential factors,
$\exp(-E_1t_y)$ and $\exp\big(-E_2(t_x-t_y)\big)$, ensure that for
large time separations, $t_y$ and $t_x-t_y$, the contributions from
the lightest states dominate. The three-point correlation function is
illustrated in the diagram
\begin{center}
\begin{picture}(160,20)(0,10)
\ArrowLine(20,20)(80,20)\Text(50,25)[b]{$H_1$}
\ArrowLine(80,20)(140,20)\Text(110,25)[b]{$H_2$}
\GCirc(20,20){1}{0}\GCirc(140,20){1}{0}
\GCirc(80,20){3}{0.5}\Text(80,14)[t]{$t_y$}\Text(80,27)[b]{$O$}
\Text(20,14)[t]{$0$}\Text(140,14)[t]{$t_x$}\Text(160,14)[t]{.}
\end{picture}
\end{center}
All the elements on the right-hand side of eq.(\ref{eq:c3states}) can
be determined from two-point correlation functions, with the exception
of the matrix element $\langle H_2|O|H_1\rangle$. Thus by computing
two- and three-point correlation functions the matrix element
$\langle H_2|O|H_1\rangle$ can be determined.

The computation of three-point correlation functions is useful, for
example, in studying semileptonic and radiative weak decays of hadrons,
e.g. if $H_1$ is a $B$-meson, $H_2$ a $D$ meson and $O$ the vector
current $\bar{b}\gamma^\mu c$, then from this correlation function we
obtain the form factors relevant for semileptonic $B\to D$ decays.

I end this brief summary of lattice computations of hadronic matrix
elements with a word about the determination of the lattice spacing
$a$.  It is conventional to introduce the parameter
$\beta=6/g_0^2(a)$, where $g_0(a)$ is the bare coupling constant in
the theory with the lattice regularization. It is $\beta$ $\big($or
equivalently $g_0(a)$$\big)$ which is the input parameter in the
simulation, and the corresponding lattice spacing is then determined
by requiring that some physical quantity (which is computed in lattice
units) is equal to the experimental value~\footnote{The bare quark
masses are also parameters which have to be determined; one has to use
as many phyical quantities as there are unknown paramters.}. For
example, one may compute $m_\rho a$, where $m_\rho$ is the mass of the
$\rho$-meson, and determine the lattice spacing $a$ by dividing the
result by $769$~MeV.

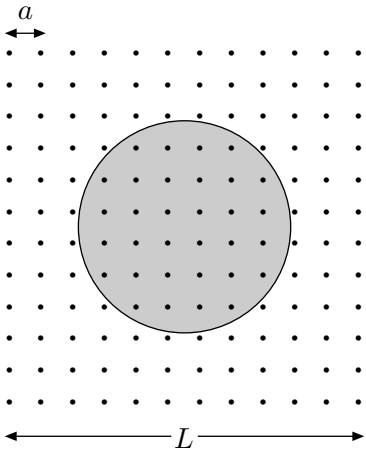
\begin{figure}[th]
\begin{center}
\begin{picture}(132,160)(0,-10)
\GCirc(66,66){40}{0.8}
\matrixput(0,0)(12,0){12}(0,12){12}{\circle*{1.5}}
\Text(6,145)[b]{$a$}
\ArrowLine(11,139)(12,139)\ArrowLine(1,139)(0,139)\Line(1,139)(11,139)
\Text(66,-10)[t]{$L$}\Line(1,-13)(61,-13)\Line(71,-13)(131,-13)
\ArrowLine(1,-13)(0,-13)\ArrowLine(131,-13)(132,-13)
\end{picture}
\caption{Schematic diagram representing a lattice containing a hadron.
$a$ and $L$ are the lattice spacing and length of the lattice respectively.} 
\label{fig:lattice}
\end{center}
\end{figure}

\subsubsection{Sources of Uncertainty in Lattice Computations:}
\label{subsubsec:errors}

Although lattice computations provide the opportunity, in principle,
to evaluate the non-perturb\-ative QCD effects in weak decays of heavy
quarks from first principles and with no model assumptions or free
parameters, in practice the precision of the results is limited by the
available computing resources. For these computations to make sense it
is necessary for the lattice to be sufficiently large to accommodate
the particle(s) being studied ($L\gg 1$\,fm say, where $L$ is the
spatial length of the lattice), and for the spacing between
neighbouring points ($a$) to be sufficiently small so that the results
are not sensitive to the granularity of the lattice ($a\lqcd\ll 1$),
see fig.~\ref{fig:lattice}.  The number of lattice points in a
simulation is limited by the available computing resources; current
simulations are performed with about 16--20 points in each spatial
direction (up to about 64 points if the effects of quark loops are
neglected, i.e.\ in the so called ``quenched'' approximation). Thus it
is possible to work on lattices which have a spatial extent of about 2
fm and a lattice spacing of 0.1 fm, perhaps satisying the above
requirements.  I now outline the main sources of uncertainty in these
computations:
\begin{itemize}
\item{\em Statistical Errors:} The functional integrals in
Eq.~(\ref{eq:vev}) are evaluated by Monte-Carlo techniques. This leads
to sampling errors, which decrease as the number of field
configurations included in the estimate of the integrals is increased.
\item{\em Discretization Errors:} These are artefacts due to the
finiteness of the lattice spacing. Much effort is being devoted to
reducing these errors either by performing simulations at several values
of the lattice spacing and extrapolating the results to the continuum
limit ($a=0$), or by ``improving'' the lattice formulation of QCD so
that the error in a simulation at a fixed value of $a$ is formally
smaller~\cite{improvement}$-$\cite{hsmpr}. In particular, it has
recently been shown to be possible to formulate lattice QCD in such a
way that the discretization errors vanish quadratically with the lattice
spacing~\cite{luscher}, even for non-zero quark masses~\cite{mrsstt}. 
This is in distinction with the traditional Wilson
formulation~\cite{wilson}, in which the errors vanish only linearly. In
some of the simulations performed in recent years  improved
actions and operators have been used. In some of these 
these studies the improvement is implemented at tree-level, so that
the artefacts
formally vanish more quickly $\big($like $a \alpha_s(a)$$\big)$ than for
the Wilson action. This tree-level improved action
is denoted as the SW (after Sheikholeslami-Wohlert who first
proposed it~\cite{sw}) or ``clover'' action.
\item{\em Extrapolations to Physical Quark Masses:} It is generally
not possible to use physical quark masses in simulations. For the
light ($u$- and $d$-) quarks the masses must be chosen such that the
corresponding $\pi$-mesons are sufficiently heavy to be insensitive to
the finite volume of the lattice. In addition, as the masses of the
quarks are reduced the computer time necessary to maintain the
required level of precision increases rapidly. For the heavy quarks
$Q$ (i.e.\ for $c$, and particularly for $b$) the masses must be
sufficiently small so that the discretization errors, which are of
$O(m_Q a)$ or $O(m_Q^2a^2)$, are small. The results obtained in the
simulations must then be extrapolated to those corresponding to
physical quark masses.
\item{\em Finite Volume Effects:} We require that the results we
obtain be independent of the size of the lattice. Thus, in principle,
the spatial size of the lattice $L$ should be $\gg 1$\,fm (in practice
$L\gtrsim 2$--3\,fm), and as mentioned above, we cannot use very light
$u$- and $d$-quarks (in order to avoid very light pions whose
correlation lengths, i.e.\ the inverses of their masses, would be of
$O(L)$ or greater).
\item{\em Contamination from Excited States:} These are the
uncertainties due to the effects of the excited states, represented by
the ellipsis in Eq.~(\ref{eq:c2exp2}). In most simulations, this is
not a serious source of error (it is, however, more significant in
computations with static quarks).  Indeed, by evaluating correlation
functions $\langle J_H(x)\,J'_H(0)\rangle$ for a variety of
interpolating operators $\{J_H, J'_H\}$, it is possible to obtain the
masses and matrix elements of the excited hadrons (for a recent
example see~\cite{excited}).
\item{\em Lattice-Continuum Matching:} The operators used in lattice
simulations are bare operators defined with the lattice spacing as the
ultra-violet cut-off. From the matrix elements of the bare lattice
composite operators, we need to obtain those of the corresponding
renormalized operators defined in some standard continuum
renormalization scheme, such as the $\overline\textrm{MS}$ scheme. The
relation between the matrix elements of lattice and continuum
composite operators involves only short-distance physics, and hence
can be obtained in perturbation theory. Experience has taught us,
however, that the coefficients in lattice perturbation theory can be
large, leading to significant uncertainties (frequently of $O(10\%)$
or more).  For this reason, non-perturbative techniques to evaluate
the renormalization constants which relate the bare lattice operators
to the renormalized ones have been developed, using chiral Ward
identities where possible~\cite{ward} or by imposing an explicit
renormalization condition~\cite{npren} (see also
Refs.~\cite{luscher,luscher2}), thus effectively removing this source
of uncertainty in many important cases.
\begin{figure}
\hbox to\hsize{\hfill
\epsfxsize0.68\hsize\epsffile{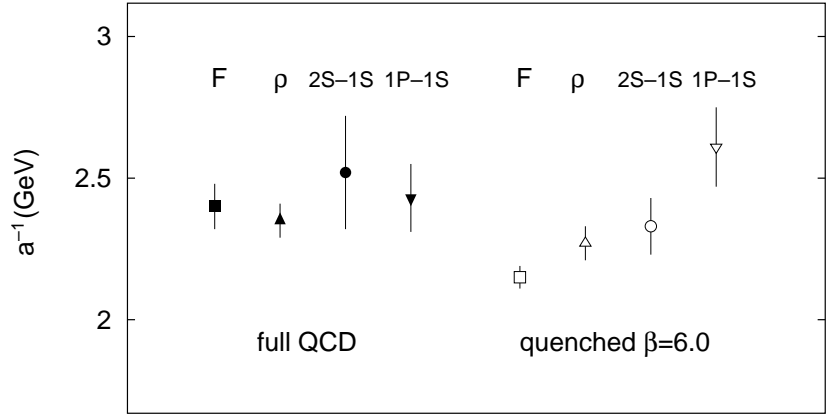}
\hfill}
\caption[]{Comparison of lattice spacings (the quantity plotted is
actually the \emph{inverse} lattice spacing, $a^{-1}$) determined from
different physical quantities in quenched and unquenched simulations
by the SESAM and T$\chi$L collaborations~\cite{henning}. $F$ denotes
the value determined from the static quark potential and $\rho$
denotes the value determined from the $\rho$-meson mass, while
$2S$--$1S$ and $1P$--$1S$ denote values obtained from energy level
splittings in $Q \bar Q$ bound states using the lattice formulation of
nonrelativistic QCD.}
\label{fig:sesamtxl}
\end{figure}
\item{\em ``Quenching'':} In most current simulations the matrix
elements are evaluated in the ``quenched'' approximation, in which the
effects of virtual quark loops are neglected. For each gluon
configuration $\{U_\mu(x)\}$, the functional integral over the quark
fields in Eq.~(\ref{eq:vev}) can be performed formally, giving the
determinant of the Dirac operator in the gluon background field
corresponding to this configuration. The numerical evaluation of this
determinant is possible, but is computationally very expensive, and
for this reason the determinant is frequently set equal to its average
value, which is equivalent to neglecting virtual quark
loops. Gradually, however, unquenched calculations are beginning to be
performed, e.g.\ in fig.~\ref{fig:sesamtxl} we show the lattice
spacing obtained by the SESAM and T$\chi$L collaborations from four
physical quantities in both quenched and unquenched
simulations~\cite{henning}. In the quenched case there is a spread of
results of about $\pm 10\%$, whereas in the unquenched case the spread
is smaller (although the errors are still sizeable for some of the
quantities used).  In the next 3--5 years it should be possible to
compute most of the physical quantities discussed below without
imposing this approximation.
\end{itemize}

\subsubsection{Lattice Calculations of $B\to\rho,\pi$ Semileptonic
Decays:}
\label{subsubsec:vublattice}

\begin{figure}
\unit 0.7\hsize
\hbox to\hsize{\hss\vbox{\offinterlineskip
\epsfxsize\unit
\epsffile[-25 54 288 232]{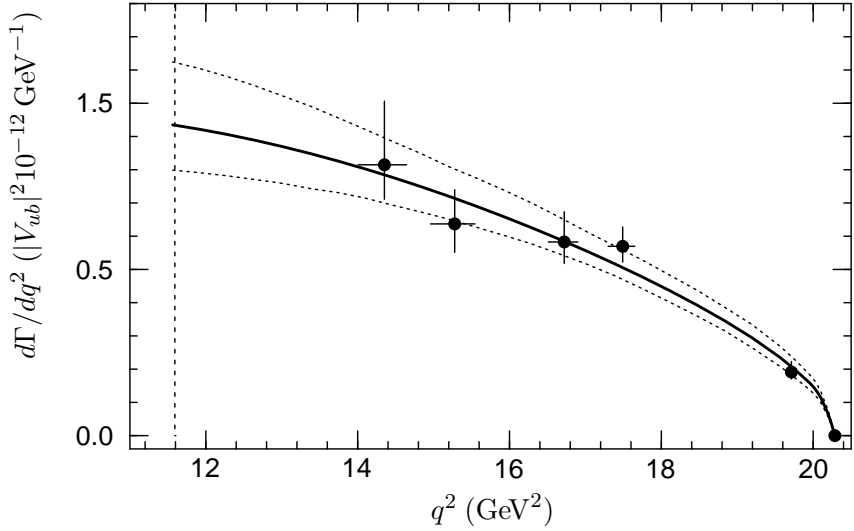}
\point 0 0.56 {\begin{sideways}
$d\Gamma/dq^2\ (|V_{ub}|^2 10^{-12}\gev^{-1})$\end{sideways}}
}\hss}
\hbox to\hsize{\hss
\hbox to\unit{\kern0.5\unit$q^2\ (\!\gev^2)$\hfill}\hss}
\caption[]{Differential decay rate as a function of $q^2$ for the
semileptonic decay $\bar B^0\to\rho^+l^-\bar\nu_l$, taken
from~\cite{ukqcd:btorho}. Points are measured lattice data, solid
curve is fit from eq.~(\ref{eq:distr2}) with parameters given in
eq.~(\ref{eq:ab-vals}). The dashed curves show the variation from the
statistical errors in the fit parameters. The vertical dotted line
marks the charm endpoint.}
\label{fig:vub}
\end{figure}

Having discussed the basics of lattice computations, I now briefly
present some results from recent simulations studying $B\to\rho,\pi$
semileptonic decays. I start by making the simple observation that from
lattice simulations we can only obtain the form factors for part of the
physical phase space for these decays.  In order to control
discretization errors we require that the three-momenta of the $B$,
$\pi$ and $\rho$ mesons be small in lattice units. This implies that we
determine the form factors at large values of momentum transfer $q^2 =
(p_B-p_{\pi,\rho})^2$. Experiments can already reconstruct exclusive
semileptonic $b\to u$ decays (see, for example, the review
in~\cite{jrp:ichep96}) and as techniques improve and new facilities
begin operation, we can expect to be able to compare the lattice form
factor calculations directly with experimental data at large $q^2$. A
proposal in this direction was made by UKQCD~\cite{ukqcd:btorho} for
$\btorho$ decays. To get some idea of the precision that might be
reached, they parametrize the differential decay rate distribution near
$\qsqmax$ by:
\begin{equation}
\frac{d\Gamma(\btorho)}{dq^2}
 =  10^{-12}\,\frac{G_F^2|V_{ub}|^2}{192\pi^3M_B^3}\,
q^2 \, \lambda^{\frac{1}{2}}(q^2)
 \, a^2\left( 1 + b(q^2{-}\qsqmax)\right),
\label{eq:distr2}
\end{equation}
where $a$ and $b$ are parameters, and the phase-space factor $\lambda$
is given by $\lambda(q^2) = (m_B^2+m_\rho^2 - q^2)^2 - 4
m_B^2m_\rho^2$. The constant $a$ plays the role of the IW function
evaluated at $\omega =1$ for heavy-to-heavy transitions, but in this case
there is no symmetry to determine its value at leading order in the
heavy quark effective theory. UKQCD obtain~\cite{ukqcd:btorho}
\begin{equation}
\begin{array}{ccc}
a = 4.6 \err{0.4}{0.3} \pm 0.6 \gev & \ \mbox{and}\ &
b = (-8 \err 46) \times 10^{-2} \gev^2.
\end{array} \label{eq:ab-vals} \end{equation}
The fits are less sensitive to $b$, so it is less well-determined. The
result for $a$ incorporates a systematic error dominated by the
uncertainty ascribed to discretization errors and would lead to an
extraction of $|V_{ub}|$ with less than 10\% statistical error and
about 12\% systematic error from the theoretical input.  The
prediction for the $d\Gamma/dq^2$ distribution based on these numbers
is presented in Fig.~\ref{fig:vub}.  With sufficient experimental data an
accurate lattice result at a single value of $q^2$ would be sufficient
to fix $|V_{ub}|$.

In principle, a similar analysis could be applied to the decay
$\btopi$. However, UKQCD find that the difficulty  of performing the
chiral extrapolation to a realistically light pion from the unphysical
pions used in the simulations makes the results less certain. 
The $B\to\pi$ decay also has a smaller fraction of events at high
$q^2$, so it will be more difficult experimentally to extract
sufficient data in this region for a detailed comparison.

We would also like to know the full $q^2$ dependence of the form
factors, which involves a large extrapolation in $q^2$ from the high
values where lattice calculations produce results, down to $q^2=0$. In
particular the important radiative decay $\btokstargamma$ (which is
dominated by penguin diagrams in the Standard Model) occurs at $q^2=0$,
so that existing lattice simulations cannot make a direct calculation of
the necessary form factors. Much effort is being devoted to
extrapolating the lattice results down to small values of $q^2$
exploiting the HQET, light-cone sum-rules~\cite{lcsr} and axiomatic
properties of quantum field theory. These studies are beyond the scope
of these lectures and I refer the students to the review
articles~\cite{fs,lp97} for a discussion and references to the original
articles.

\section{Lecture 3: $K^0-\bar K^0$ and $B^0-\bar B^0$ Mixing}
\label{sec:mixing}

In this lecture I will discuss $K$-$\bar K$ and $B$-$\bar B$ mixing in
turn, and consider the implications of the experimental measurements of
these processes for our understanding of the unitarity triangle.

\subsection{$K^0$-$\bar K^0$ Mixing, $\epsilon_K$ and
$\epsilon^\prime/\epsilon$} 
\label{subsec:epsilon}

The process of $K^0$-$\bar K^0$ mixing appears in the Standard Model at
one-loop level through the box-diagrams of fig.~\ref{fig:box} (with the
$b$-quark replaced by the $s$-quark), together with possible
long-distance contributions. The $CP$-eigenstates ($K_1$ and $K_2$) are
linear combinations of the two strong-interaction
eigenstates~\footnote{I use the phase convention so that $CP|K^0\rangle
=|\bar K^0\rangle$\, .}:
\begin{equation}
\kone = \frac{1}{\sqrt{2}}\,(\kzero + \kzerobar)\hspace{0.7in}
CP\kone = \kone
\label{eq:k1}\end{equation}
and
\begin{equation}
\ktwo = \frac{1}{\sqrt{2}}\,(\kzero - \kzerobar)\hspace{0.7in}
CP\ktwo = -\ktwo\ .
\label{eq:k2}\end{equation}
Because of the complex phase in the CKM-matrix, the physical states (the mass
eigenstates) differ from $\kone$ and $\ktwo$ by a small admixture of the other
state:
\begin{equation}
\ks=\frac{\kone +
\bar\epsilon\,\ktwo}{(1+|\bar\epsilon|^2)^{\frac{1}{2}}} \ \ \
\textrm{and}
\ \ \ \kl=\frac{\ktwo +
\bar\epsilon\,\kone}{(1+|\bar\epsilon|^2)^{\frac{1}{2}}}\ ,
\label{eq:kskldef}\end{equation}
(the parameter $\bar\epsilon$ depends on the phase convention chosen for
$\kzero$ and $\kzerobar$).

For exclusive decays of $K$-mesons (which have angular momentum zero)
into two or three pion states, the two pion states are $CP$-even and the
three-pion states are $CP$-odd. This implies that the dominant decays are:
\begin{equation}
K_S\to\pi\pi\ \ \ \textrm{and}\ \ \ K_L\to 3\pi\ ,
\label{eq:dominant}\end{equation}
which is the reason why $K_L$ is much longer lived than $K_S$ ($L$ and
$S$ stand for \emph{long} and \emph{short} respectively). $K_L$ and
$K_S$ are not $CP$-eigenstates, however, and the decay $K_L\to 2\pi$ may
occur as a result of the small component of $\kone$ in $|K_L\rangle$,
and similarly the  decay $K_S\to 3\pi$ may occur because of the small
component of $\ktwo$ in $|K_S\rangle$. Such $CP$-violating decays which
occur due to the fact that the mass eigenstates are not $CP$-eigenstates
are called \emph{indirect $CP$-violating decays}. As mentioned above,
the parameter $\bar\epsilon$ depends on the convention used to define
the phases of the states. A measure of the strength of indirect
$CP$-violation is given by the physical parameter $\epsilon_K$ defined
by the ratio:
\begin{equation}
\epsilon_K\equiv\frac{A\left(K_L\to(\pi\pi)_{I=0}\right)}
{A\left(K_S\to(\pi\pi)_{I=0}\right)}=(2.280\pm0.013)\,10^{-3}\,
e^{i\frac{\pi}{4}}\ ,
\label{eq:epskdef}\end{equation}
where $A$ represents the amplitude for the corresponding process and
$I=0$ implies that the two pions are in an isospin 0 state. The
numerical value in eq.~(\ref{eq:epskdef}) is the empirical result.

Directly $CP$-violating decays are those in which a $CP$-even (-odd)
state decays into a $CP$-odd (-even) one:
\begin{equation}
K_L\ \ \propto \ \ K_2\ \  + \ \ \bar\epsilon K_1\ .
\label{eq:dirindir}\end{equation}\nopagebreak
\begin{picture}(300,124)(-27,-10)
\Line(194,120)(194,80)\ArrowLine(194,80)(194,78)
\Text(194,73)[t]{$\pi\pi$}\Text(186,90)[rb]{Direct $(\epsilon^\prime)$}
\Line(237,120)(237,100)\Line(237,100)(257,100)\ArrowLine(257,100)(259,100)
\Text(264,100)[l]{$\pi\pi$}\Text(235,90)[tl]{Indirect $(\epsilon_K)$}
\end{picture}

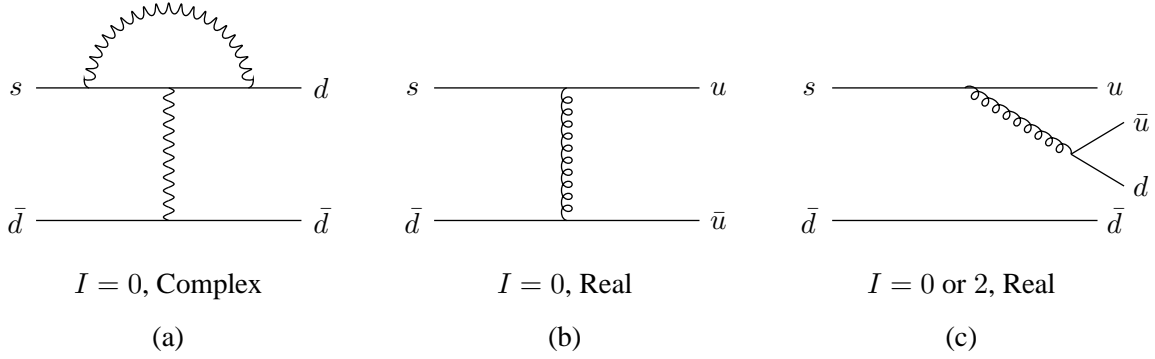
\begin{figure}[ht]
\begin{center}
\begin{picture}(420,120)(0,-30)
\Line(10,10)(110,10)\Line(10,60)(110,60)\Photon(60,10)(60,60){2}{10}
\GlueArc(60,60)(30,180,0){2}{20}
\Text(5,60)[r]{$s$}\Text(5,10)[r]{$\bar d$}
\Text(115,60)[l]{$d$}\Text(115,10)[l]{$\bar d$}
\Text(60,-10)[t]{$I=0$, Complex}
\Text(60,-30)[t]{(a)}
\Line(160,10)(260,10)\Line(160,60)(260,60)\Gluon(210,10)(210,60){2}{10}
\Text(155,60)[r]{$s$}\Text(155,10)[r]{$\bar d$}
\Text(265,60)[l]{$u$}\Text(265,10)[l]{$\bar u$}
\Text(210,-10)[t]{$I=0$, Real}
\Text(210,-30)[t]{(b)}
\Line(310,10)(410,10)\Line(310,60)(410,60)
\Gluon(360,60)(400,35){2}{9}
\Text(305,60)[r]{$s$}\Text(305,10)[r]{$\bar d$}
\Line(400,35)(420,47)\Line(400,35)(420,23)
\Text(415,60)[l]{$u$}\Text(425,47)[l]{$\bar u$}
\Text(425,23)[l]{$d$}
\Text(360,-10)[t]{$I=0$ or $2$, Real}
\Text(360,-30)[t]{(c)}\Text(415,10)[l]{$\bar d$}
\end{picture}
\end{center}
\caption{Three subprocesses contributing to $K\to\pi\pi$ decays.}
\label{fig:direct}\end{figure}

\vspace{-70pt}\noindent Consider the possible quark subprocesses
contributing to $K\to\pi\pi$ decays shown in fig.~\ref{fig:direct}. The
diagrams in fig.~\ref{fig:direct}(b) and (c) are purely real, whereas
that of fig.~\ref{fig:direct}(a) is complex. The two pions in the final
state can have total isospin $I=0$ or 2; $I=1$ is not possible by Bose
symmetry. The final states in the diagrams of fig.~\ref{fig:direct}(a)
and (b) clearly have $I=0$, whereas that in the diagram of
fig.~\ref{fig:direct}(c) can have either $I=0$ or $I=2$. Thus direct
$CP$-violation in kaon decays manifests itself as a non-zero relative
phase between the $I=0$ and $I=2$ amplitudes.

Of course the gluonic corrections to the diagrams in
fig.~\ref{fig:direct} generate \emph{strong phases}, which we denote by
$\delta_{0,2}$, where the suffix $0$ or $2$ represents the isospin of
the final state. These phases are independent of the form of the weak
Hamiltonian. Using the Clebsch-Gordan coefficients for the combination
of two isospin 1 pions we write:
\begin{eqnarray}
A(K^0\to\pi^+\pi^-)& = &\sqrt{\frac{2}{3}}\,A_0\,e^{i\delta_0} +
\sqrt{\frac{1}{3}}\,A_2\,e^{i\delta_2}\label{eq:charged}\\
A(K^0\to\pi^0\pi^0)& = &\sqrt{\frac{2}{3}}\,A_0\,e^{i\delta_0} -
2\sqrt{\frac{1}{3}}\,A_2\,e^{i\delta_2}\label{eq:neutral}\ .
\end{eqnarray}
The parameter $\epsilon^\prime$, which is used as a measure of direct
$CP$-violation in kaon decays, is defined by:
\begin{equation}
\epsilon^\prime = \frac{\omega}{\sqrt{2}}\,e^{i\phi}\,
\left(\frac{\textrm{Im}\, A_2}{\textrm{Re}\, A_2} -
\frac{\textrm{Im}\, A_0}{\textrm{Re}\, A_0}\right)\ , 
\label{eq:epsprimedef} \end{equation}
where~\footnote{$\epsilon^\prime$ is manifestly zero if the phases of the
$I=0$ and $I=2$ amplitudes are the same.}
\begin{equation}
\omega\equiv\frac{\textrm{Re}\,A_2}{\textrm{Re}\,A_0}\ \ \ \textrm{and}
\ \ \ \phi=\frac{\pi}{2}+\delta_2-\delta_0\simeq\frac{\pi}{4}\ .
\label{eq:omegadef}\end{equation}

Experimentally the two parameters $\epsilon_K$ (which, following
standard conventions I rename from now on as $\epsilon$,
$\epsilon\equiv\epsilon_K$) and $\epsilon^\prime$ can be determined by
measuring the ratios:
\begin{eqnarray}
\eta_{00}&\equiv&\frac{A(K_L\to\pi^0\pi^0)}{A(K_S\to\pi^0\pi^0)}\,\simeq
\ \epsilon - 2\epsilon^\prime\label{eq:etazz}\\ 
\eta_{+-}&\equiv&\frac{A(K_L\to\pi^+\pi^-)}{A(K_S\to\pi^+\pi^-)}\,\simeq
\ \epsilon + \epsilon^\prime\ .
\label{eq:etapm}\end{eqnarray}
Direct $CP$-violation is found to be considerable smaller than indirect
violation. By measuring the decays and using
\begin{equation}
\left|\frac{\eta_{00}}{\eta_{+-}}\right|^2\simeq 1\,-\, 6\,\textrm{Re}
\,\left(\frac{\epsilon^\prime}{\epsilon}\right)\ +\cdots\ ,
\label{eq:etazzetapm}\end{equation}
the NA31~\cite{na31} and E371~\cite{e371} experiments find $(23\pm
7)\,10^{-4}$ and $(7.4\pm 5.9)\,10^{-4}$ respectively for
$\epsilon^\prime/\epsilon$. The Particle Data Group~\cite{pdg}
summarises these results as:
\begin{equation}
\epsilon^\prime/\epsilon=(1.5\pm 0.8)\,10^{-3}\ .
\label{eq:epsilonpepsilonno}\end{equation}
We will have to wait for the results of the current generation of
experiments, which will have a sensitivity of $O(10^{-4})$, to confirm
or deny whether the value is indeed different from zero.

\subsubsection{Determination of $\epsilon$}
\label{subsubsec:epsilon}

In this subsection I discuss the theoretical determination of
$\epsilon$, following the discussion and notation in the review article
by Buras and Fleischer~\cite{bf}. We need to know the matrix
element:
\begin{equation}
\langle\bar K^0\,|\,\mathcal{H}^{\Delta
S=2}_{\scriptstyle{eff}}\,|\,K^0\rangle\ ,
\label{eq:epsme}\end{equation}
where we write the effective Hamiltonian in the form:
\begin{eqnarray}
\mathcal{H}^{\Delta S=2}_{\scriptstyle{eff}}& = &
\frac{G_F^2}{16\pi^2}M_W^2\left[\lambda_c^2\eta_1S_0(x_c)\ +\
\lambda_t^2\eta_2S_0(x_t) + 2
\lambda_c\lambda_t\eta_3S_0(x_c,x_t)\right]\nonumber\\
& &\mbox{\ \ \ }\times 
\left[\alpha_s^{(3)}(\mu)\right]^{-\frac{2}{9}}\,\left[1 +
\frac{\alpha_s^{(3)}(\mu)}{4\pi}J_3\right]\,O^{\Delta S=2}(\mu)\ +\
\textrm{h.c.}\ ,
\label{eq:heffexpr}\end{eqnarray}
the renormalization scale $\mu$ is chosen to be much smaller than
$m_c$, $x_i = m_i^2/M_W^2$ and  $\lambda_i=V_{id}V_{is}^*$ (the
dependence on $\lambda_u$ is eliminated by unitarity, $x_u=0$). The
non-perturbative QCD effects for $K^0$-$\bar K^0$ mixing are all contained
in the matrix elements of the single local composite operator
\begin{equation}
O^{\Delta S=2}(\mu) = \bar s\gamma^\mu(1-\gamma^5)d\
\bar s\gamma_\mu(1-\gamma^5)d\ ,
\label{eq:ods2def}\end{equation}
the remaining terms in eq.~(\ref{eq:heffexpr}) can be calculated in
perturbation theory. The $S_0$'s are the expressions for the box-diagrams
without QCD corrections:
\begin{eqnarray}
S_0(x) & = & \frac{4x-11x^2+x^3}{4(1-x)^2}\,-\,\frac{3x^3\ln(x)}{2(1-x)^3}
\label{eq:sox}\\
S_0(x_c,x_t) & = & x_c\left[ \ln\left(\frac{x_t}{x_c}\right) -
\frac{3x_t}{4(1-x_t)}-\frac{3x_t^2\ln(x_t)}{4(1-x_t)^2}\right] + \cdots\ .
\label{eq:soxx}\end{eqnarray}
$\eta_1,\eta_2$ and $\eta_3$ are mass-dependent short-distance QCD factors,
at NLO (next-to leading order)
\begin{equation}
\eta_1=1.38\pm 0.20\ \ \ \eta_2=0.57\pm 0.01\ \ \textrm{and}\ \
\eta_3=0.47\pm 0.04\ ,
\label{eq:etas}\end{equation}
where the errors reflect the uncertainty in $\lqcd$ and the quark
masses. The remaining factors in eq.~(\ref{eq:heffexpr}) are
short-distance QCD effects, whose $\mu$-dependence cancels that in the
operator $O^{\Delta S=2}$ (in the NDR-renormalization scheme
$J_3$=1.895).

As mentioned above, all the non-perturbative QCD corrections are contained
in the matrix elements of $O^{\Delta S=2}$. It is conventional to introduce
the $B_K$ parameter by the definition:
\begin{equation}
\langle\bar K^0\,|\,\bar s\gamma^\mu(1-\gamma^5)d\ \bar
s\gamma_\mu(1-\gamma^5)d\,|K^0\rangle
\equiv\frac{8}{3}m_K^2f_K^2B_K(\mu)\ .
\label{eq:bkdef}\end{equation}
The motivation for introducing the parameter $B_K$ in this way comes
from the vacuum saturation approximation (in which $B_K=1$), but there
is no loss of generality in this definition. $B_K(\mu)$ depends on the
renormalization scale, and it is convenient to introduce an (almost)
renormalization group invariant $B_K$ by:
\begin{equation}
B_K=B_K(\mu)\left[\alpha_s^{(3)}(\mu)\right]^{-\frac{2}{9}}\,\left[1 +
\frac{\alpha_s^{(3)}(\mu)}{4\pi}J_3\right]\ .
\end{equation}
The weakest link in the theoretical calculation of $\epsilon$ is the
evaluation of $B_K$. Many lattice computations have been performed, and
a recent compilation of results gave $B_K = 0.90\pm
0.06$~\cite{flynnwarsaw}. A calculation based on the $1/N$ approximation
gave $B_K=0.70\pm 0.10$~\cite{bk1overn}. 
Buras and Fleischer perform their analysis with
\begin{equation}
B_K=0.75\pm 0.15\ .
\label{eq:bkbest}\end{equation}

$\epsilon$ is given in terms of the quantities introduced above by
\begin{equation}
\epsilon = C_\epsilon\,B_K\,\textrm{Im}\,\lambda_t\left\{
\textrm{Re}\,\lambda_c\left[\eta_1S_0(x_c)-\eta_3S_0(x_c,x_t)\right]
-\textrm{Re}\,\lambda_t\,\eta_2S_0(x_t)\right\}\,e^{i\frac{\pi}{4}}\ ,
\label{eq:epsexpr}\end{equation}
where
\begin{equation}
C_\epsilon = \frac{G_F^2 f_K^2 m_K M_W^2}{6\sqrt{2}\pi^2(\Delta M_K)}
=3.78\cdot 10^4
\label{eq:ceps}\end{equation}
and the mass difference $\Delta M_K = M_{K_L}-M_{K_S}$.

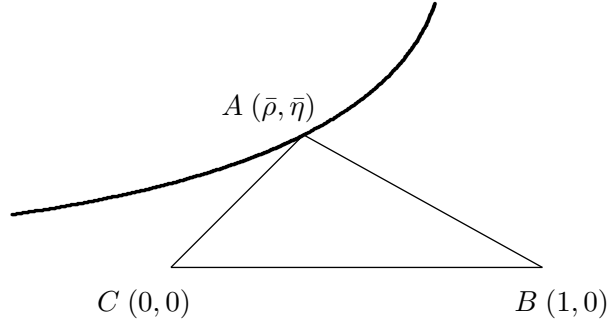
\begin{figure}[ht]
\begin{center}
\begin{picture}(210,100)(-50,0)
\Line(10,10)(150,10)\Line(10,10)(60,60)\Line(60,60)(150,10)
\Text(65,65)[rb]{$A\ (\bar\rho,\bar\eta)$}
\Text(140,2)[lt]{$B\ (1,0)$}
\Text(0,2)[t]{$C\ (0,0)$}
\linethickness{0.4mm}
\curve(-50,30,60,60,110,110)
\end{picture}
\end{center}
\caption{A precise determination of $\epsilon$ would fix
the vertex $A$ to lie on a a hyperbola (schematically
represented by the solid curve).}
\label{fig:eps}\end{figure}

We obtain information about the unitarity triangle by comparing the
theoretical prediction with the experimentally measured value of
$\epsilon$. In terms of $\bar\eta$ and $\bar\rho$ the prediction is
the hyperbola 
\begin{equation}
\bar\eta\left[
(1-\bar\rho)A^2\eta_2S_0(x_t)+P_0\right]\,A^2\,B_K\,=\,0.226\ ,
\label{eq:bkhyperbola}\end{equation}
where $P_0=0.31\pm0.02$. This is illustrated in fig.~\ref{fig:eps}.

\subsubsection{The $\Delta I = 1/2$ Rule and $\epsilon^\prime/\epsilon$}
\label{subsubsec:deltai12}

In this subsection I briefly review the elements in the theoretical
predictions for $\epsilon^\prime/\epsilon$ and related quantities. I
start by rewriting eq.~(\ref{eq:epsprimedef}) in the form:
\begin{equation}
\epsilon^\prime = - \frac{\omega}{\sqrt{2}}\,\xi\,(1-\Omega)\,e^{i\phi}\ ,
\end{equation}
where
\begin{equation}
\omega=\frac{\textrm{Re}\,A_2}{\textrm{Re}\,A_0}\, ,
\ \ \xi=\frac{\textrm{Im}\,A_0}{\textrm{Re}\,A_0}\, ,
\ \ \Omega = \frac{1}{\omega}\frac{\textrm{Im}\,A_2}{\textrm{Im}\,A_0}\
\ \textrm{and}\ \ 
\phi=\frac{\pi}{2}+\delta_2-\delta_0\simeq\frac{\pi}{4}\ .
\label{eq:parsdef}\end{equation}

Even after about 40 years, we still do not understand theoretically the
$\Delta I = 1/2$ rule, i.e. the empirical observation that $\omega\simeq
1/22$. The rate for transitions in which the isospin changes by 1/2
($\Delta I = 1/2$ transitions) is enhanced by a factor of about 450 over
those in which it changes by 3/2 ($\Delta I = 3/2$ transitions). There
is a relatively small enhancement of the $\Delta I = 1/2$ amplitudes, by
a factor of 2-3, due to perturbative effects (i.e. in the calculation of
the Wilson coefficient function). The remaining enhancement is assumed
to be due to long-distance QCD effects, perhaps because the
matrix-elements of penguin operators are large, but this has still to be
convincingly demonstrated.

In the recent analyses of $\epsilon^\prime/\epsilon$, one combines the
experimental values for the real parts of the amplitudes
(Re\,$A_0=3.33\cdot 10^{-7}\,$GeV, Re\,$A_2=1.50\cdot 10^{-8}\,$GeV,
$\omega=0.045$), with theoretical predictions for the imaginary parts. 
The effective Hamiltonian for these $\Delta S=1$ transitions takes the
form:
\begin{equation}
\mathcal{H}_{eff}(\Delta S=1)=\frac{G_F}{\sqrt{2}}\,
\sum_{i=1}^{10}\left(z_i(\mu) + \tau y_i(\mu)\right)\,O_i(\mu)\ ,
\label{eq:heffdseq1}\end{equation}
where $\tau=-V_{td}V^*_{ts}/V_{ud}V^*_{us}$. There are 10 independent
operators in the $\Delta S=1$ effective Hamiltonian, and the major
uncertainty is in the values of their matrix elements and in the mass of
the strange quark $m_s$. The most complete analyses have been performed by
Buras et al~\cite{bjl} who find:
\begin{equation}
(-1.2\cdot10^{-4})<\epsilon^\prime/\epsilon<16\cdot10^{-4}\ \ (I)\ \ ;
\ \ \epsilon^\prime/\epsilon = (3.6\pm 3.4)\,10^{-4}\ \ (II)\ ,
\label{eq:epspepsbf}\end{equation}
where $I$ and $II$ refer to two different ways of treating the
many uncertainties (more and less conservative respectively), and by
the Rome group~\cite{epsprome} who find
\begin{equation}
\epsilon^\prime/\epsilon = (4.6\pm 3.0)\,10^{-4}\ .
\label{eq:epspepsrome}\end{equation}
The various contributions to $\epsilon^\prime/\epsilon$ have different
signs, and there are significant cancellations. For this reason even the
expected sign cannot be predicted with full confidence. 

I end this subsection with a brief summary of the main points:
\begin{itemize}
\item An understanding of the $\Delta I = 1/2$ rule would be an
important milestone in controlling non-perturbative QCD effects.
This is a realistic, but non-trivial, challenge for the lattice
community.
\item A measurement of a non-zero value for $\epsilon^\prime/\epsilon$
would be a very important qualitative step in particle physics;
confirming the existence of direct $CP$-violation.
\item The uncertainties in the values of the matrix elements, make it
difficult to make a precise prediction for $\epsilon^\prime/\epsilon$.
We expect it to be at the several $\times 10^{-4}$ level, but accidental
cancellations between contributions with opposite signs may make it
smaller than this. In the coming years, experiments at CERN, FNAL and at
Da$\phi$ne will be sensitive to a value of about $10^{-4}$.
\end{itemize}

\subsection{$B^0$-$\bar B^0$ Mixing}
\label{subsec:bbar}

In this subsection we consider $B$-$\bar B$ mixing, from which we get
information about the $V_{td}$ and $V_{ts}$ elements of the CKM-matrix. 
I start with some general formalism for the mixing of a neutral
psudoscalar meson $P^{\,0}$ with its antiparticle $\bar P^{\,0}$. I
write the wave-function in two-component form:
\begin{equation}
|\psi(t)\rangle = \left(\begin{array}{c}
a(t) \\ b(t)
\end{array}\right)
\equiv a(t)\,|P^0\rangle + b(t)\,|\bar P^0\rangle\ .
\label{eq:wf}\end{equation}
The time dependence of $|\psi(t)\rangle$ is given by the Schr\" odinger
equation:
\begin{equation}
i\frac{d}{dt}\,|\psi(t)\rangle = \left( M - \frac{i\Gamma}{2}\right)
|\psi(t)\rangle\ ,
\label{eq:se}\end{equation}
where $M-i\Gamma/2$ is the $2\times 2$ ``mass-matrix", whose elements
are equal to $\langle P_i^0|\mathcal{H}_{eff}|P_j^0\rangle /2M_P$. Using
the optical theorem, the absorbtive part of this matrix comes from real
intermediate states:
\begin{equation}
\Gamma_{ij}=\frac{1}{2M_P}\sum_n\langle P^{\,0}_i\,|
\mathcal{H}_{W}\,|n\rangle\langle n\,|\mathcal{H}_{W}\,|P^{\,0}_j
\rangle\, 2\pi\,\delta(E_n-M_P)\ .
\label{eq:gammaij}\end{equation}
For $B^0$-$\bar B^0$ mixing, there are no contributions to $\Gamma_{12}$ 
from intermediate states containing the top quark.

The mass-matrix takes the form
\begin{equation}
M-\frac{i\Gamma}{2} = \left(
\begin{array}{cc}
A & \ \, p^2 \\ q^2 & \ \, A
\end{array}\right)\ ,
\label{eq:massform}\end{equation}
where the diagonal terms are equal by CPT-invariance, and $A$, $p^2$ and
$q^2$ are complex. The eigenvalues of this matrix are $M_{1,2}=A\pm pq$,
so that the mass difference between the two physical eigenstates is
\begin{equation}
\Delta M = M_1 - M_2 = 2pq = 2 (M_{12}-i\Gamma_{12})^{\frac{1}{2}}
(M_{12}^*-i\Gamma_{12}^*)^{\frac{1}{2}}\ .
\end{equation}

For the $B$-system, the box diagrams have contributions from $t$-quarks
in the loops (see fig.~\ref{fig:box}). By unitarity these contribute
only to the real part of the mass-matrix. The mass behaviour of the box
diagrams in fig.~\ref{fig:box}, and the large value of $V_{tb}$ imply
that these contributions dominate and that $|\Gamma_{12}|\ll |M_{12}|$
for $B$-$\bar B$ mixing (this is not the case for $K$-$\bar K$ mixing).
Thus $\Delta M \simeq 2 |M_{12}|$, and can be calculated from the box
diagrams.

The evaluation of the box diagrams follows similarly to that in the kaon
system. The non-perturbative QCD effects are contained in the matrix
element of the $\Delta B=2$ operator:
\begin{equation}
O^{\Delta B=2} = \bar b\gamma^\mu (1-\gamma^5)d\
\bar b\gamma_\mu (1-\gamma^5)d\ ,
\end{equation}
and it is also convenient and conventional to introduce the $B_B$
parameter analogously to the definition of $B_K$ in
eq.~(\ref{eq:bkdef}). The theoretical expression for the mass difference
is:
\begin{equation}
\Delta M=\frac{G_F^2}{6\pi^2}\,\eta_B\,M_B\,B_B\,f_B^2\,M_W^2\,S_0(x_t)\,
|V_{td}|^2\ ,
\end{equation}
where $\eta_B$ is a pertubative QCD correction ($\eta_B=0.55\pm0.01$).
Throughout this discussion I have been implicitly assuming that we are
considering neutral $B_d$ mesons, but an analogous discussion holds for
the $B_s$ system. In the kaon system, all the non-perturbative QCD
effects were contained in the parameter $B_K$, since the leptonic decay
constant $f_K$ is a measured quantity. For the $B$-mesons this is not
the case, and one is obliged to use model estimates or lattice or
sum-rule calclulations for both the decay constants $f_B$ and the
$B_B$-parameters (or of the matrix element of $O^{\Delta B=2}$ itself).
For example, in a recent compilation of lattice results we
obtained~\cite{fs}:
\begin{equation}
f_{B_d}=(170\pm 35)\gev,\ \ f_{B_s}=(195\pm 35)\gev,\ \
B_{B_d}=1.4(1),
\label{eq:bresults1}\end{equation}
and
\begin{equation}
\xi\equiv\frac{f_{B_s}\sqrt{B_{B_s}}}
{f_{B_d}\sqrt{B_{B_d}}}=1.14(8)\ .
\label{eq:bresults}\end{equation}

Combining all the elements one obtains:
\begin{eqnarray}
\Delta M_{B_d} & = & 0.50\,\textrm{ps}^{-1}\,\left[
\frac{f_{B_d}\sqrt{B_{B_d}}}{200\mev}\right]^2\,
\left(\frac{\bar m_t(m_t)}{170\gev}\right)^{1.52}\,
\frac{|V_{td}|}{0.0088}\ \frac{\eta_B}{0.55}
\label{eq:dmd}\\
\Delta M_{B_s} & = & 15.1\,\textrm{ps}^{-1}\,\left[
\frac{f_{B_s}\sqrt{B_{B_s}}}{240\mev}\right]^2\,
\left(\frac{\bar m_t(m_t)}{170\gev}\right)^{1.52}\,
\frac{|V_{ts}|}{0.040}\ \frac{\eta_B}{0.55}\ .
\label{eq:dms}\end{eqnarray}

The mass difference $\Delta M$ is a measure of the oscillation frequency
to change from a $B^0$ to a $\bar B^0$ and vice-versa. Imagine that we
start with a $B^0$ at time $t=0$, then the propabilities that we have a
$B^0$ or $\bar B^0$ at a later time $t$ are $\exp(-\Gamma
t)\cos^2(\Delta M\,t/2)$ and $\exp(-\Gamma t)\sin^2(\Delta M\,t/2)$
respectively. Defining $P(B^0)$ and $P(\bar B^0)$ to be the probabilities
(obtained by integrating over time) that when the meson decays it is
a $B$ or $\bar B$ respectively, we obtain
\begin{equation}
P(B^0) = \frac{1}{2}\,\left[\frac{1}{\Gamma}+\frac{\Gamma}{\Gamma^2+
(\Delta M)^2}\right]\ \ \textrm{and}\ \
P(\bar B^0) = \frac{1}{2}\,\left[\frac{1}{\Gamma}-\frac{\Gamma}{\Gamma^2+
(\Delta M)^2}\right]\ .
\end{equation}
The ratio of these integrated probabilities is given by
\begin{equation}
r\equiv\frac{P(B^0)}{P(\bar B^0)}\ =\ \frac{x^2}{2+x^2}\ ,
\label{eq:rdef}\end{equation}
where $x=\Delta M/\Gamma$. The 1996 Particle Data Group review quotes:
\begin{equation}
\Delta M_{B_d} = (0.474 \pm 0.031)\,\textrm{ps}^{-1}\ \ \textrm{and}\ \
x_d = 0.73\pm 0.05\ ,
\label{eq:bdresults}\end{equation}
in agreement with theoretical expectations and
\begin{equation}
\Delta M_{B_s} > 5.9\,\textrm{ps}^{-1}\ \ \textrm{and}\ \
x_s > 9.5\ \ \textrm{at }95\%\ \textrm{CL}\ .
\label{eq:bsresults}\end{equation}
The oscillation in the $B_s$ system are too rapid for observation in
experiments which have been performed up to date.

\begin{figure}[ht]
\begin{center}
\begin{picture}(210,120)(0,0)
\Line(10,10)(150,10)\Line(10,10)(60,60)\Line(60,60)(150,10)
\Text(65,65)[rb]{$A\ (\bar\rho,\bar\eta)$}
\Text(140,2)[lt]{$B\ (1,0)$}
\Text(0,2)[t]{$C\ (0,0)$}
\SetWidth{1.5}
\CArc(150,10)(103,0,170)
\end{picture}
\end{center}
\caption{A precise determination of $|V_{td}/V_{cb}|$ from studies of
$B-\bar B$ mixing would fix the vertex $A$ to lie on a circle centered on
the vertex $B$ (schematically represented by the solid curve).}
\label{fig:vtd}\end{figure}
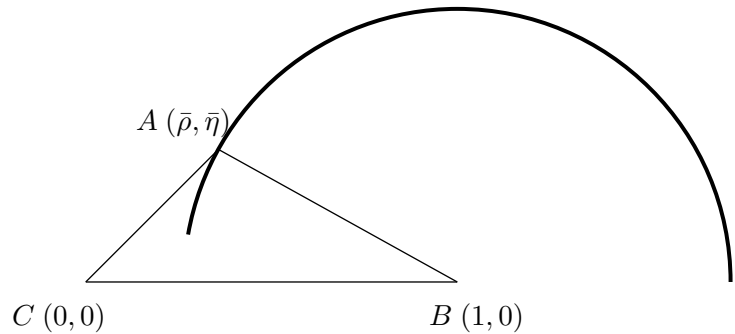

What do these measurements imply for the determination of the position of the
unitarity triangle? In principle a measurement of $\Delta M_{B_d}$ allows for
a determination of the $V_{td}$ element of the CKM matrix. The side $BA$ of
the unitarity triangle is given by:
\begin{equation}
\frac{|V_{td}V_{tb}^*|}{|V_{cd}V_{cb}^*|}=\frac{1}{\lambda}
\frac{|V_{td}|}{|V_{cb}|}=\sqrt{(1-\bar\rho)^2+\bar\eta^2}\ ,
\label{eq:vtdut}\end{equation}
so that a measurement of $V_{td}$ (assuming knowledge of $V_{cb}$) would
imply that the locus of possible positions of the vertex $A$ lies on a
given circle centred on $B$, see fig.~\ref{fig:vtd}.

\section{Lecture 4: CP-Violation in $B$-Decays and Inclusive $B$-Decays}
\label{sec:cpinc}

In this final lecture I will consider two topics in $B$-physics,
$CP$-violation (and mixing-induced $CP$-violation, in particular) and
the application of the heavy quark expansion to inclusive decays (and to
beauty-lifetimes and the semileptonic branching ratio, in particular). I
will also make some brief remarks about how difficult it is to make
predictions for exclusive nonleptonic decays.

\subsection{$CP$-Violation in $B$-Decays}
\label{subsec:cpvb}

Measurements of $CP$-asymmetries in neutral $B$-meson decays into
$CP$-eigenstates will allow us to address 3 fundamental questions:
\begin{enumerate}
\item[i)] Is the phase in the CKM matrix the only source of
$CP$-violation?
\item[ii)] What are the precise values of the CKM parameters? We shall
see that the principal systematic errors, particularly those due to
hadronic uncertainties, are reduced significantly in some cases.
\item[iii)] Is there new physics in the quark-sector?
\end{enumerate}

A particularly important class of decays of neutral $B$-meson systems,
are \emph{mixing induced} $CP$-violating decays (these are not possible
in decays of charged $B$-mesons). I now briefly review this important
topic, for which  we need to return to the formalism introduced in the
discussion of $B$-$\bar B$ mixing in subsec.~\ref{subsec:bbar} above.
The two neutral mass-eigenstates are given by
\begin{equation}
|B_L\rangle = \frac{1}{\sqrt{p^2 + q^2}}\,\left(p\,|B^0\rangle + q
\,|\bar B^0\rangle\right)\ \ \textrm{and}\ \
|B_H\rangle = \frac{1}{\sqrt{p^2 + q^2}}\,\left(p\,|B^0\rangle - q
\,|\bar B^0\rangle\right)\ .
\label{eq:bhldef}\end{equation}
Starting with a $B^0$ meson at time $t=0$, its subsequent evolution is
governed by the Schr\" odinger equation (\ref{eq:se}):
\begin{equation}
|B^0_{\mathrm{phys}}(t)\rangle = g_+(t)\,|B^0\rangle +
\left(\frac{q}{p}\right) g_-(t)|\bar B^0\rangle\ ,
\end{equation}
where
\begin{eqnarray}
g_+(t) & = & \exp\left[-\frac{\Gamma t}{2}\right]\,\exp[-iMt]\,
\cos\left(\frac{\Delta M\,t}{2}\right)\ ,\label{eq:gp}\\ 
g_-(t) & = & \exp\left[-\frac{\Gamma t}{2}\right]\,\exp[-iMt]\,
i\sin\left(\frac{\Delta M\,t}{2}\right)\ ,\label{eq:gm} 
\end{eqnarray}
and $M=(M_H+M_L)/2$~\footnote{Starting with a $\bar B^0$ meson at $t=0$,
the time evolution is $|\bar B^0_{\mathrm{phys}}(t)\rangle = (p/q)
\,g_-(t)|\bar B^0\rangle + g_+(t)\,|\bar B^0\rangle$.}.

\subsubsection{Decays of Neutral $B$-Mesons into $CP$-Eigenstates}
\label{subsubsec:cpvneutral}

Let $f_{CP}$ be a $CP$-eigenstate and $A,\bar A$ be the amplitudes
\begin{equation}
A\equiv\langle f_{CP}|\mathcal{H}|B^0\rangle\ \ \textrm{and}\ \ 
\bar A\equiv\langle f_{CP}|\mathcal{H}|\bar B^0\rangle\ .
\label{eq:aabardef}\end{equation}
Defining
\begin{equation}
\lambda\equiv\frac{q}{p}\,\frac{\bar A}{A}
\label{eq:lambdadef}\end{equation}
we have
\begin{equation}
\langle f_{CP}\,|\,\mathcal{H}\,|\,B^0_{\mathrm{phys}}\rangle  = 
A\,[g_+(t)\,+\,\lambda\,g_-(t)]\ \ \ \textrm{and}\ \ \ 
\langle f_{CP}\,|\,\mathcal{H}\,|\,\bar B^0_{\mathrm{phys}}\rangle = 
A\,\frac{p}{q}\,[g_-(t)\,+\,\lambda\,g_+(t)]\ .
\label{eq:btof}\end{equation}
The time-dependent rates for initially pure $B^0$ or $\bar B^0$ states
to decay into the $CP$-eigenstate $f_{CP}$ at time $t$ are given by:
\begin{eqnarray}
\Gamma(B^0_{\mathrm{phys}}(t)\to f_{CP}) & = & |A|^2\,e^{-\Gamma t}
\left[\frac{1+|\lambda|^2}{2}\,+\,\frac{1-|\lambda|^2}{2}\cos(\Delta M\,t)
-\textrm{Im}\,\lambda\,\sin(\Delta M\,t)\right]\label{eq:btofcp}\\
\Gamma(\bar B^0_{\mathrm{phys}}(t)\to f_{CP}) & = & |A|^2\,e^{-\Gamma t}
\left[\frac{1-|\lambda|^2}{2}\,+\,\frac{1-|\lambda|^2}{2}\cos(\Delta M\,t)
+\textrm{Im}\,\lambda\,\sin(\Delta M\,t)\right]\ .\label{eq:bbartofcp}
\end{eqnarray}
The time-dependent asymmetry is defined as:
\begin{eqnarray}
\mathcal{A}_{f_{CP}}(t)& \equiv& \frac{\Gamma(B^0_{\mathrm{phys}}(t)\to
f_{CP}) - \Gamma(\bar B^0_{\mathrm{phys}}(t)\to f_{CP})}
{\Gamma(B^0_{\mathrm{phys}}(t)\to
f_{CP}) + \Gamma(\bar B^0_{\mathrm{phys}}(t)\to f_{CP})}
\label{eq:asymdef}\\ 
& = & \frac{(1-|\lambda|^2)\cos(\Delta M\,t) - 2\textrm{Im}\,\lambda\,
\sin(\Delta M\,t)}{1+|\lambda|^2}\ . 
\label{eq:aexpr}\end{eqnarray}
If, as is the case in some important applications, $|q/p|=1$ (which is
the case if $\Delta\Gamma<<\Delta M$) and $|\bar A/A|=1$ (examples of
this will be presented below), then $|\lambda|=1$ and the first term on
the right-hand side of eq.(\ref{eq:aexpr}) vanishes.

The form of the amplitudes $A$ and $\bar A$ is:
\begin{equation}
A=\sum_{i}\,A_i\,e^{i\delta_i}\,e^{i\phi_i}\ \ \ \textrm{and}\ \ \ 
\bar A=\sum_{i}\,A_i\,e^{i\delta_i}\,e^{-i\phi_i}
\label{eq:ampsexpr}\end{equation}
where the sum is over all the contributions to the process, the $A_i$
are real, $\delta_i$ are the strong phases and the $\phi_i$ are the
phases from the CKM matrix. In the most favourable situation, all the
contributions have a single CKM phase ($\phi_D$ say) and
$\bar A/A=\exp(-2i\phi_D)$. Under the assumption we are making that
$\Gamma_{12}<<M_{12}$, $q/p=\sqrt{M_{12}^*/M_{12}}\equiv\exp(-2i\phi_M)$,
and $\lambda=\exp(-2i(\phi_D+\phi_M))$. Thus
\begin{equation}
\textrm{Im}\,\lambda = -\sin(2(\phi_D+\phi_M))\ .
\label{eq:iml}\end{equation}

From the box diagrams of fig.~\ref{fig:box} we obtain:
\begin{equation}
\left(\frac{q}{p}\right)_{B_d}=\frac{V_{td}V_{tb}^*}{V^*_{td}V_{tb}}
\ \ \ \textrm{and}\ \ \
\left(\frac{q}{p}\right)_{B_s}=\frac{V_{ts}V_{tb}^*}{V^*_{ts}V_{tb}}\ .
\label{eq:qoverpb}\end{equation}

To illustrate the above discussion let us consider three processes in which the
$b$-quark decays through the subprocess $b\to d_ju_i\bar u_i$. The
corresponding tree-level diagram is 
\begin{equation}
\mbox{\begin{picture}(80,50)(80,20)
\Line(10,45)(80,45)\Gluon(40,45)(80,15){2}{10}
\Line(80,15)(105,25)\Line(80,15)(105,5)
\Text(5,45)[r]{$b$}\Text(85,45)[l]{$u_i$}
\Text(110,25)[l]{$d_j$}\Text(110,5)[l]{$\bar u_i$}
\end{picture}}
\textrm{for which}\ \ \ \ \frac{\bar A}{A}=\frac{V_{ib}V_{ij}^*}{V_{ib}^*V_{ij}}
\ .\label{eq:abarovera}\end{equation}

\vspace{25pt}
\paragraph{$\mathbf{B_d\to J/\Psi\,K_S:}$}
In this case
\begin{equation}
\lambda(B\to J/\Psi K_S) = \frac{V_{td}V_{tb}^*}{V_{td}^*V_{tb}}\ 
\frac{V_{cs}V_{cd}^*}{V_{cs}^*V_{cd}}\ 
\frac{V_{cb}V_{cs}^*}{V_{cb}^*V_{cs}}\ =-\sin(2\beta)\ .
\label{eq:li}\end{equation}
The first factor in eq.~(\ref{eq:li}) is the factor $(q/p)_{B_d}$ in
eq.~(\ref{eq:qoverpb}), the third factor is the analogous one for the
final state kaon, and the second factor is $\bar A/A$ as in
eq.~(\ref{eq:abarovera}) with $u_i=c$ and $b_j=s$. We recall from the
discussion of the first lecture that the angle
\begin{equation}
\beta=\arg\left(-\frac{V_{cd}V_{cb}^*}{V_{td}V_{tb}^*}\right)\ .
\label{eq:anglebeta}\end{equation}

There is also a small penguin contribution to this process, due to the
subprocess:
\begin{center}
\begin{picture}(140,60)(0,20)
\Line(10,50)(130,50)\Text(5,50)[r]{$b$}\Text(135,50)[l]{$s$}
\GlueArc(70,50)(30,180,0){3}{20}
\Text(65,56)[b]{$t$}\Text(100,75)[b]{$W$}
\Photon(70,50)(120,20){2}{10}
\Line(120,20)(130,28)\Line(120,20)(130,12)
\Text(95,27)[t]{$\mathcal{G}$}\Text(140,12)[c]{,}
\end{picture}
\end{center}
which is proportional to $V_{tb}V_{ts}^*$, whose phase is, to a very
good approximation, equal to that of $V_{cb}V_{cs}^*$. Thus hadronic
uncertainties are negligible in the determination of the angle $\beta$
from this process, and for this reason we can consider it a
\emph{gold-plated} one. This is an (almost) ideal situation.

\paragraph{$\mathbf{B_d\to\pi^+\pi^-:}$}
 
For this process, in the notation of eq.~(\ref{eq:abarovera}), we
have $u_i=u$ and $d_j=d$, so that
\begin{equation}
\frac{\bar A}{A}=\frac{V_{ub}V_{ud}^*}{V_{ub}V_{ud}^*}
\label{eq:btopipi}\end{equation}
and Im\,$\lambda=\sin(2\alpha)$, where from the first lecture
we recall that
\begin{equation}
\alpha=\arg\left(-\frac{V_{td}V_{tb}^*}{V_{ud}V_{ub}^*}\right)\ .
\label{eq:anglealpha}\end{equation}

In this case there is also a small penguin contribution, which is
proportional to $V_{td}^*V_{tb}$, which has a different phase from the
tree diagram. This implies that the hadronic uncertainites are greater
than for the process $B_d\to J/\Psi K_S$, although they are still
reasonably small (and can be reduced further using isospin analysis).

\paragraph{$\mathbf{B_s\to\rho^0 K_S:}$}

For this process in the notation of eq.~(\ref{eq:abarovera}), we
have $u_i=u$ and $d_j=d$ again, so that 
\begin{equation}
\lambda(B_s\to \rho K_S) = \frac{V_{ts}V_{tb}^*}{V_{ts}^*V_{tb}}\ 
\frac{V_{ub}V_{ud}^*}{V_{ub}^*V_{ud}}\ 
\frac{V_{cd}V_{cs}^*}{V_{cd}^*V_{cs}}\ .
\label{eq:liii}\end{equation}
Using the unitarity relations and the fact that
\begin{equation}
\gamma=\arg\left(-\frac{V_{ud}V_{ub}^*}{V_{cd}V_{cb}^*}\right)\ ,
\label{eq:anglegamma}\end{equation}
we find that Im\,$\lambda=-\sin(2\gamma)$.

\begin{figure}[ht]
\begin{center}
\begin{picture}(300,150)(0,0)
\Line(30,10)(250,10)
\Line(30,10)(270,120)\Line(250,10)(270,120)
\Line(30,10)(100,80)\Line(250,10)(100,80)
\Text(140,5)[t]{$A(B^+\to\bar D^0K^+)=A(B^-\to D^0K^-)$}
\Text(55,45)[r]{$\sqrt{2}\,A(B^+\to D^0_+K^+)$}
\Text(265,65)[l]{$A(B^-\to \bar D^0K^-)$}
\Text(280,130)[br]{$\sqrt{2}\,A(B^-\to D^0_+K^-)$}
\Line(212,128)(232,108)\ArrowLine(232,108)(234,106)
\Text(70,85)[lb]{$A(B^+\to \bar D^0K^+)$}
\Line(130,84)(122,76)\ArrowLine(122,76)(120,74)
\Text(248,27)[r]{$2\gamma$}\CArc(250,10)(32,79,156)
\end{picture}
\end{center}
\caption{Six amplitudes from which the angle $\gamma$ can (in principle
at least) be determined.}
\label{fig:triangles}\end{figure}
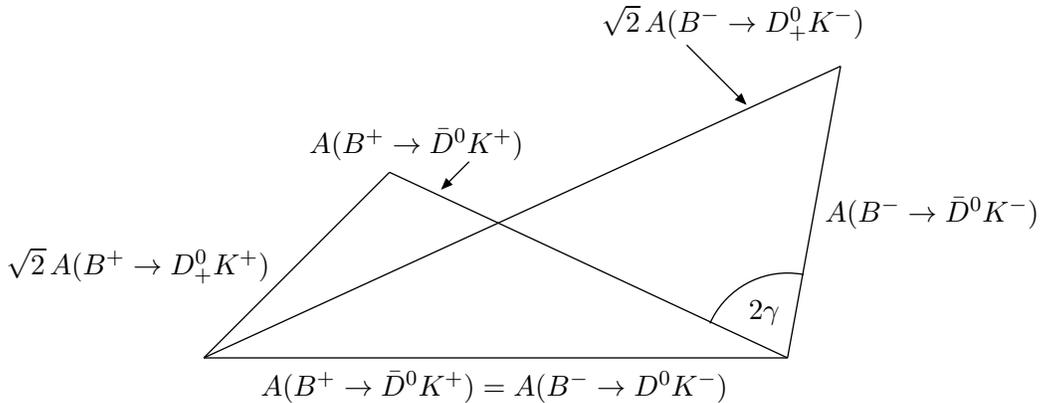

Although this process is frequently proposed as a potential way of
measuring the angle $\gamma$, the penguin contributions are relatively
large and Buras and Fleischer stress that it is a \emph{wrong
way}~\cite{bf}. It has been suggested that it might be possible to use
combinations of $B\to DK$ amplitudes to determine the angle
$\gamma$~\cite{gw}. This is
illustrated in fig.~\ref{fig:triangles}, where
$|D^0_+\rangle=1/\sqrt{2}(\,|D^0\rangle + |\bar D^0\rangle\,)$, but
clearly a measurement of all the amplitudes is a huge experimental
challenge.

The exciting experimental and theoretical program of research into
mixing induced $CP$-asymetries will continue at the (near) future
$B$-factories and dedicated $B$-experiments. 

\subsubsection{$CP$-Violation in Charged $B$-Decays}
\label{subsubsec:cpvcharged}

Observation of $CP$-violation in decays of charged $B$-mesons would
demonstrate the existence of direct $CP$-violation, but the hadronic
uncertainties make it unlikely that they can be used to determine the
parameters of the unitarity triangle with the required precision. Since
we need an interference between different contributions to the
amplitude, we write the amplitudes in the form:
\begin{equation}
A(B^-\to f)=V_1\,A_1\,e^{i\delta_1} + V_2\,A_2\,e^{i\delta_2}
\ \ \ \textrm{and}\ \ \
A(B^+\to \bar f)=V_1^*\,A_1\,e^{i\delta_1} + V_2^*\,A_2\,e^{i\delta_2}
\ ,\label{eq:bchargedtof}\end{equation}
where I have exhibited two contributions to each amplitude, $V_{1,2}$
are CKM-matrix elements and the $\delta_{1,2}$ are strong interaction
phases. For example in the process $B^+\to\pi^0K^+$ the interference is
between curent-current and penguin contributions, and in the decay
$B^+\to\bar K^0K^+$ it is between penguin diagrams with internal $u$ and
$c$ quarks.

The $CP$-violating asymmetry is now defined by
\begin{equation}
\mathcal{A}_{CP}\equiv\frac{\Gamma(B^+\to\bar f){-}\Gamma(B^-\to f)}
{\Gamma(B^+\to\bar f) + \Gamma(B^-\to f)}
=\frac{2\,\textrm{Im}\,(V_1V_2^*)\,\sin(\delta_1{-}\delta_2)\,A_1A_2}
{|V_1|^2A_1^2 + |V_2|^2A_2^2  + 2\textrm{Re}\,(V_1V_2^*)\cos(\delta_1{-}
\delta_2)\,A_1A_2}\ ,
\label{eq:acpcharged}\end{equation}
from which we see that in order to have a non-zero asymmetry we require
both Im\,$(V_1V_2^*)$ and $\sin(\delta_1-\delta_2)$ to be non-zero. The
strong interaction phases are very difficult to quantify, so it is
hard to use measurements of these $CP$-asymetries to determine the CKM
matrix elements.

\subsection{Inclusive Nonleptonic Decays of Beauty Hadrons}
\label{subsec:nlinclusive}

In this subsection I discuss inclusive nonleptonic decays of beauty hadrons
in general, and two very interesting phenomenological problems in particular
the lifetimes of the hadrons and their semileptonic
branching ratios. The discussion will use the formalism of Bigi \etal
(see ref.~\cite{bbsuv} and references therein), developed and used by
them and many other groups, in which inclusive quantities are expanded
in inverse powers of the mass of the heavy quark, e.g.
\begin{equation}
\Gamma(H_b) = \frac{G_F^2m_b^5|V_{cb}|^2}{192\pi^3}
\left\{c_3\left(1+ \frac{\lambda_1+3\lambda_2}{2m_b^2}\right)
+c_5\frac{\lambda_2}{m_b^2}+O\left(\frac{1}{m_b^3}\right)\right\}\ ,
\label{eq:widthexpansion}\end{equation}
where $\Gamma$ is the full or partial width of a beauty hadron $H_b$,
$c_{3,5}$ are coefficients which can be computed in perturbation theory
(the coefficient functions obtained when matching QCD onto the HQET)
and $\lambda_{1,2}$ are the matrix elements of the kinetic energy 
and chromomagnetic operators respectively:
\begin{equation}
\lambda_1(B)=\frac{1}{2m_{H_b}}\langle H_b|\,\bar h(i\vec D)^2 h\,|H_b
\rangle\ \ \ \ \textrm{and}\ \ \ \ 
\lambda_2  =  \frac{1}{3}\frac{\langle H_b|\bar
h\frac{1}{2}\sigma_{ij}G^{ij}h|H_b\rangle}{2m_{H_b}} \ ,
\label{eq:l12def}\end{equation}
and $h$ is the field of the heavy (static) quark. An important feature
of the general expression in eq.~(\ref{eq:widthexpansion}) is the
absence of terms of $O(1/m_b)$, which is a consequence of the absence of
any operators of dimension 4 which can appear in the corresponding
OPE~\cite{nooneoverm}.

I will not discuss the derivation of the expansion in
eq.~(\ref{eq:widthexpansion}) in detail. One starts, similarly to the
derivation of the OPE for moments of deep-inelastic structure functions,
by using the optical theorem which states that the width is given by the
imaginary part of the forward elastic amplitude. In this case this
amplitude has two insertions of the weak Hamiltonian. Since the
$b$-quark is heavy and decays into lighter states, the separation of
these two insertions is small, and eq.~(\ref{eq:widthexpansion}) is the
corresponding short distance expansion (for the structure functions it
is a light-cone expansion). We now apply this expansion to a study of
beauty lifetimes and the semileptonic brabching ratio.

\subsubsection{Beauty Lifetimes}
\label{subsubsec:lifetimes}

Using the expression in eq.~(\ref{eq:widthexpansion}) for the widths, one
readily finds the following results for the ratios of lifetimes:
\begin{eqnarray}
\frac{\tau(B^-)}{\tau(B^0)} & = & 1 + O\left(\frac{1}{m_b^3}\right)
\label{eq:ratiomesons}\\
\frac{\tau(\Lambda_b)}{\tau(B^0)} & = &
1 + \frac{\mu_\pi^2(\Lambda_b)-\mu_\pi^2(B)}{2m_b^2} +
c_G \frac{\mu_G^2(\Lambda_b)-\mu_G^2(B)}{m_b^2}
+ O\left(\frac{1}{m_b^3}\right)\nonumber\\ 
& = & (0.98\pm0.01) + O\left(\frac{1}{m_b^3}\right)\ ,
\label{eq:ratiolambdab}\end{eqnarray}
where $\mu_\pi^2 = -\lambda_1$ and $\mu_G^2=3\lambda_2$.
In order to obtain the result in eq.(\ref{eq:ratiolambdab}), one needs
to know the difference of the kinetic energies of the $b$-quark in the
baryon and meson. To leading order in the heavy quark expansion we have:
\begin{equation}
\mu_\pi^2(\Lambda_b)-\mu_\pi^2(B) = -\frac{M_BM_D}{2}
\left(\frac{M_{\Lambda_b} - M_{\Lambda_c}}{M_B-M_D} - \frac{3}{4}
\frac{M_{B^*}- M_{D^*}}{M_B - M_D}-\frac{1}{4}\right)\ .
\label{eq:kediffs}\end{equation}
From equation~(\ref{eq:kediffs}), and using the recent measurement of
$m_{\Lambda_B}$ from CDF~\cite{pdg2}, one finds that the right hand side
is very small (less than about 0.01 GeV$^2$). The matrix elements of the
chromomagnetic operator are obtained from the mass difference of the
$B^*$- and $B$-mesons (see eq.(\ref{eq:l12def})\,) and from the fact that
the two valence quarks in the $\Lambda_b$ are in a spin-zero state
(which implies that $\mu_G^2(\Lambda_b)=0$). The
theoretical predictions in eqs.~(\ref{eq:ratiomesons}) and
(\ref{eq:ratiolambdab}) can be compared with the experimental
measurements
\begin{equation}
\begin{array}{ccc}
\displaystyle
\frac{\tau(B^-)}{\tau(B^0)}  =  1.06 \pm 0.04 &\ \mbox{and}\ &
\displaystyle
\frac{\tau(\Lambda_b)}{\tau(B^0)}  =  0.79\pm 0.05\ .
\label{eq:ratiolambdabexp}\end{array}\end{equation}
The discrepancy between the theoretical and experimental results
for the ratio $\tau(\Lambda_b)/\tau(B^0)$ in
eqs.~(\ref{eq:ratiolambdab}) and (\ref{eq:ratiolambdabexp}) is
notable. It raises the question of whether the $O(1/m_b^3)$
contributions are surprisingly large, or whether there is a more
fundamental problem. I postpone consideration of the latter
possibility and start with a discussion of the $O(1/m_b^3)$ terms.

\begin{figure}[ht]
\begin{center}
\begin{picture}(300,50)(-30,15)
\ArrowLine(10,41)(40,41)\ArrowLine(100,41)(40,41)\ArrowLine(100,41)(130,41)
\Oval(70,41)(15,30)(0)
\GCirc(40,41){1.5}{0}\GCirc(100,41){1.5}{0}
\ArrowLine(69,56)(71,56)\ArrowLine(69,26)(71,26)
\Text(20,36)[tl]{$b$}\Text(115,36)[tr]{$b$}
\Text(67,63)[bl]{$c$}\Text(67,19)[tl]{$d$}\Text(80,44)[br]{$\bar u$}
\SetWidth{2}\Line(140,41)(158,41)\ArrowLine(158,41)(160,41)
\SetWidth{0.5}\Line(180,41)(230,41)\GCirc(205,41){2}{0}
\Text(205,33)[t]{$\bar bb$}
\end{picture}
\begin{picture}(300,50)(260,40)
\ArrowLine(300,41)(330,41)\ArrowLine(330,41)(300,26)
\ArrowLine(330,41)(390,41)\Curve{(330,41)(360,56)(390,41)}
\ArrowLine(390,41)(420,41)\ArrowLine(420,26)(390,41)
\ArrowLine(361,56)(359,56)
\GCirc(330,41){1.5}{0}\GCirc(390,41){1.5}{0}
\SetWidth{2}\Line(430,41)(448,41)\ArrowLine(448,41)(450,41)
\SetWidth{0.5}\Line(475,51)(495,41)\Line(495,41)(515,51)
\Line(495,41)(475,31)\Line(515,31)(495,41)
\GCirc(495,41){2}{0}
\Text(310,46)[bl]{$b$}\Text(312,28)[tl]{$\bar d$}
\Text(361,63)[b]{$\bar u$}\Text(361,34)[t]{$c$}
\Text(410,46)[br]{$b$}\Text(407,28)[tr]{$\bar d$}
\Text(500,23)[t]{$(\bar b\Gamma q)\,(\bar q\Gamma b)$}
\end{picture}
\vspace{0.4in}
\caption{Examples of diagrams whose imaginary parts contribute to
the total rates for the decays of beauty hadrons (left-hand sides)
and the operators they correspond to in the Operator Product Expansions.
$\Gamma$ represents a Dirac matrix.\label{fig:diagrams}}
\end{center}\end{figure}
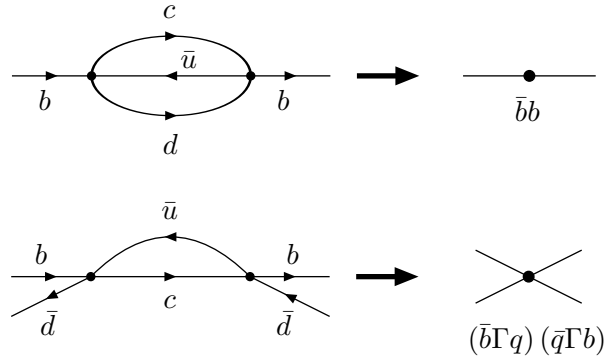

At first sight it seems strange to consider the $1/m_b^3$ corrections to
be a potential source of large corrections, when the $O(1/m_b^2)$ terms
are only about 2\%. However, it is only at this order that the
``spectator'' quark contributes, and so these contributions lead
directly to differences in lifetimes for hadrons with different light
quark constituents (consider for example the lower diagram in
Fig.~\ref{fig:diagrams}, for which, using the short-distance expansion,
one obtains operators of dimension 6).  Moreover, the coefficient
functions of these operators are relatively large, which may be
attributed to the fact that the lower diagram in Fig.~\ref{fig:diagrams}
is a one-loop graph, whereas the corresponding diagrams for the leading
contributions are two-loop graphs (see, for example, the upper diagram
of Fig.~\ref{fig:diagrams}). The corresponding phase-space enhancement
factor is 16$\pi^2$ or so. We will therefore only consider the
contributions from the corresponding four-quark operators, neglecting
other $O(1/m_b^3)$ corrections which do not have the phase space
enhancement~\cite{ns}.  For each light-quark flavour $q$, there are four
of these~\footnote{I use the notation of ref.~\cite{ns}.}:
\begin{eqnarray}
O_1\equiv \bar b\gamma_\mu(1-\gamma^5)q\, \bar q\gamma^\mu(1-\gamma^5)b
\hspace{0.32in} &;&
O_2\equiv \bar b (1+\gamma^5)q\, \bar q(1+\gamma^5)b
\label{eq:odefs}\\
T_1\equiv \bar b\gamma_\mu(1-\gamma^5)T^aq \,\bar q\gamma^\mu(1-\gamma^5)T^a b
&;&  
T_2\equiv \bar b (1+\gamma^5)T^a q \,\bar q (1+\gamma^5) T^a b
\end{eqnarray}
where $T^a$ are the generators of colour $SU(3)$ and $q$ represents the
fields of the light quarks. Thus we need to evaluate the matrix elements
of these four operators.

For mesons, following ref.~\cite{ns}, I introduce the parametrization
\begin{equation}
\begin{array}{ccc}
\langle B| O_i|B\rangle|_{\mu = m_b} \equiv
B_if_B^2M_B^2 & ; &
\langle B| T_i|B\rangle|_{\mu = m_b}  \equiv
\epsilon_if_B^2M_B^2\ ,
\end{array}
\label{eq:biepsilonidef}\end{equation} 
where $\mu$ is the renormalization scale. We have chosen to use $m_b$
as the renormalization scale. Bigi et al. (see ref.~\cite{bbsuv} and
references therein) prefer to use a typical hadronic scale, and
estimate the matrix elements using a factorization hypothesis at this
low scale. Operators renormalized at different scales can be related
using renormalization group equations in the HQET (sometimes called
hybrid renormalization~\cite{hybrid}). For example, if we assume that
factorisation holds at a low scale $\mu$ such that
$\alpha_s(\mu^2)=1/2$, then, using the (leading order) renormalization
group equations, one finds $B_1=B_2=1.01$ and
$\epsilon_1=\epsilon_2=-0.05$~\footnote{By factorization we mean that
if the $B_i$'s and $\epsilon_i$'s had been defined at this scale
(instead of $m_b$) they would have been 1 and 0 respectively.}\,. In
the limit of a large number of colours $N_c$, $B_i=O(N_c^0)$ whereas
$\epsilon_i=O(1/N_c)$.

For the $\Lambda_b$, heavy quark symmetry implies that
\begin{equation}
\begin{array}{ccc}
\displaystyle
\langle\Lambda_b|O_2|\Lambda_b\rangle=
-\frac{1}{2}\langle\Lambda_b|O_1|\Lambda_b\rangle
&\ \mbox{and}\ &\displaystyle
\langle\Lambda_b|T_2|\Lambda_b\rangle=
-\frac{1}{2}\langle\Lambda_b|T_1|\Lambda_b\rangle\ ,
\label{eq:lbhqs}\end{array}\end{equation}
so that there are only two independent parameters. It is convenient to
replace the operator $T_1$, by $\tilde O_1$ defined by
\begin{equation}
\tilde O_1 \equiv \bar b^i\gamma_\mu(1-\gamma^5) q^j\ 
\bar q^j\gamma^\mu(1-\gamma^5) b^i\ ,
\end{equation}
where $i,j$ are colour labels,
and to express physical quantities in terms of the two parameters
$\tilde B$ and $r$ defined by
\begin{eqnarray}
\langle\Lambda_b|\tilde O_1|\Lambda_b\rangle_{\mu=m_b} & \equiv &
-\tilde B \langle\Lambda_b|O_1|\Lambda_b\rangle_{\mu=m_b}
\label{eq:btildedef}\\
\frac{1}{2M_{\Lambda_b}}\langle\Lambda_b|O_1|\Lambda_b\rangle_{\mu=m_b}
& \equiv & -\frac{f_B^2M_B}{48}\,r\ .
\label{eq:rdef2}\end{eqnarray}
We do not know the values of these parameters. In quark models $\tilde B=1$,
and $r=0.2$--$0.5$. Using experimental values of the hyperfine splittings and
quark models, it
has been suggested that $r$ may be larger~\cite{rosner}, e.g.
\begin{equation}
\begin{array}{lcl}
r\simeq\frac{4}{3}\frac{M_{\Sigma_c^*}^2 - M_{\Sigma_c}^2}
{M_{D^*}^2 - M_{D}^2} = 0.9 \pm 0.1\label{eq:rc}\ \ 
& \textrm{and} &
\ \ r\simeq\frac{4}{3}\frac{M_{\Sigma_b^*}^2 - M_{\Sigma_b}^2}
{M_{B^*}^2 - M_{B}^2}  =  1.8 \pm 0.5\label{eq:rb}\ .
\end{array}\end{equation}

The lifetime ratios can now be written in terms of the six parameters
$B_{1,2},\epsilon_{1,2},\tilde B$ and $r$ (as well as $f_B$):
\begin{eqnarray}
\frac{\tau(B^-)}{\tau(B^0)} & = & 1 + \left(\frac{f_B}{200\mev}\right)^2
\left\{0.02B_1 + 0.00 B_2 -0.70 \epsilon_1 + 0.20\epsilon_2\right\}
\label{eq:mesonsresult}\\
\frac{\Lambda_b}{\tau(B^0)} & = & 0.98 + \left(\frac{f_B}{200\mev}\right)^2
\{-0.00B_1 + 0.00 B_2 -0.17 \epsilon_1 + 0.20\epsilon_2\nonumber\\ 
& & \hspace{0.6in} 
+ (-0.01 -0.02\tilde B)\,r\}\label{eq:baryonsresult}\ ,
\end{eqnarray}
where the effective weak Lagrangian has been renormalized at $\mu=m_b$.
The central question is whether it is possible, with ``reasonable"
values of the parameters, to obtain agreement with the experimental
numbers in eq.~(\ref{eq:ratiolambdabexp}). At this stage in our
knowledge, the answer depends somewhat on what is meant by
\emph{reasonable}. For example, Neubert, guided by the arguments
outlined above,  has considered these ratios by varying the parameters
in the following ranges~\cite{mnhonolulu}:
\begin{equation}
B_i,\tilde B\in\left[\frac{2}{3},\frac{4}{3}\right]\,;\ \
\epsilon_i\in\left[-\frac{1}{3},\frac{1}{3}\right]\, ;\ \ r\in[0.25, 2.5]\,;
\ \ \left(\frac{f_B}{200\mev}\right)^2\in[0.8,1.2]\ .
\label{eq:ranges}\end{equation}
He concludes that, within these ranges, it is just possible to obtain
agreement at the two standard deviation level for large values of $r$
($r\ge 1.2$) and negative values of $\epsilon_2$. Lattice studies of the
corresponding matrix elements are underway; a recent  QCD sum-rule
calculation has found a small value of $r$, 
$r\simeq 0.1$--$0.3$~\cite{colangelo}.

If the lattice calculations confirm that the parameter $r$ is small, or
find that the other parameters are not in the appropriate ranges, then
we have a breakdown of our understanding. If no explanation can be found
within the standard formulation, then we will be forced to take
seriously the possible breakdown of local duality. This is beginning to
be studied in toy field theories~\cite{blok,nardulli}.

\subsubsection{The ``Baffling" Semileptonic Branching Ratio}
\label{subsubsec:bsl}

\begin{figure}[t]
\hbox to\hsize{\hss\vbox{\offinterlineskip
\unit=0.6\hsize
\epsfxsize=\unit\epsffile{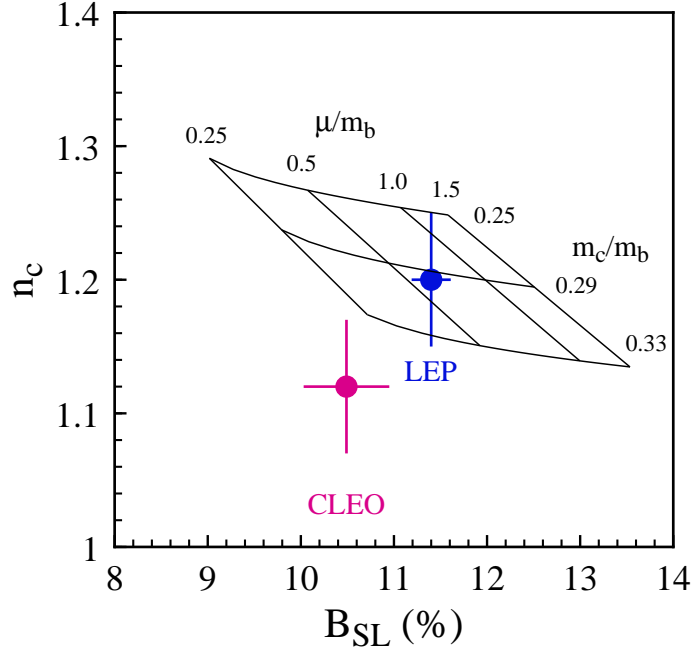}
}\hss}
\caption[]{Theoretical Prediction (shaded region) 
of the semileptonic branching ratio and charm counting. The
data points are the experimental results from high-energy
(LEP) and low energy (i.e. at the $\Upsilon(4S)$ from CLEO)
experiments.}
\label{fig:bsl}
\end{figure}

This was the name given by Blok \etal~\cite{baffling} to the
observation that the experimental value of the semileptonic branching
ratio 
\begin{equation}
B_{SL} = \frac{\Gamma(B\to Xe\bar\nu)}{\sum_{l}\Gamma(B\to Xl\bar\nu)
+ \Gamma_{had} + \Gamma_{rare}}
\label{eq:bsldef}\end{equation}
appeared to be lower than expected theoretically. In
eq.~(\ref{eq:bsldef}) the sum is over the three species of lepton, and
$\Gamma_{had}$ and $\Gamma_{rare}$ are the widths of the hadronic and
rare decays respectively. Bigi \etal\ concluded that a branching ratio
of less than 12.5\% cannot be accommodated by theory~\cite{baffling}.
Since then Bagan \etal\  have completed the calculation of the
$O(\alpha_s)$ corrections, and in particular of the $b\to c\bar cs$
component of the decay (including the effects of the mass of the charm
quark)~\cite{bagan}; these have the effect of decreasing $B_{SL}$. 
With M.~Neubert, we used this input to reevaluate the branching
ratio and charm counting ($n_c$, the average number of charmed particles
per $B$-decay)~\cite{ns} finding, e.g.
\begin{equation}
\begin{array}{ll}
B_{SL} = 12.0\pm 1.0\%\ (\mu=m_b)\ \ & 
\ \ n_c = 1.20 \mp 0.06\ \ (\mu=m_b)\label{eq:bslncmb}\\ 
B_{SL} = 10.9\pm 1.0\%\ (\mu=m_b/2)\ \ &
\ \ n_c = 1.21 \mp 0.06\ \ (\mu=m_b/2)\label{eq:bslnchalfmb}\ .
\end{array}\end{equation}
$\mu$ is the renormalization scale and the dependence on this scale is a
reflection of our ignorance of higher order perturbative corrections.
The experimental situation is somewhat confused, see Fig.~\ref{fig:bsl}.
In his compilation at the ICHEP conference last year~\cite{richman}, Richman
found that the semileptonic branching ratio obtained from $B$-mesons
from the $\Upsilon(4S)$  is~\footnote{Note that the rapporteur at the
1997 EPS conference argued that the branching ratio had been
overestimated by the LEP collaboration~\cite{feindt}.}:
\begin{equation}
B_{SL}(B)=(10.23\pm0.39)\%\ ,
\end{equation}
whereas that from LEP is:
\begin{equation}
B_{SL}(b)=(10.95\pm0.32)\%\ .
\end{equation}
The label $b$ for the LEP measurement indicates that 
the decays from beauty hadrons other than the $B$-meson are included.
Using the measured fractions
of the different hadrons and their lifetimes, and assuming that
the semileptonic widths of all the beauty hadrons are the same,
one finds:
\begin{equation}
B_{SL}(b)=(10.95\pm 0.32)\% \Rightarrow B_{SL}(B)=(11.23\pm0.34)\%\ , 
\end{equation}
amplifying the discrepancy. It is very difficult to understand such a
discrepancy theoretically, since the theoretical calculation only
involves $\Gamma_{SL}$ (and not $\Gamma_{had}$ for which the
uncertainties are much larger). In view of the experimental discrepancy,
I consider the problem of the lifetime ratio
$\tau(\Lambda_b)/\tau(B^0)$, described in
subsection~\ref{subsubsec:lifetimes} above, to be the more significant
theoretical one.

\subsubsection{Exclusive Decays of $B$-Mesons}
\label{subsubsec:exclusive}

A large amount of experimental data is becoming available, particularly
from the CLEO collaboration (see ref.~\cite{drell} and references
therein) on two-body (exclusive) nonleptonic decays of $B$-mesons. This
is an exciting new field of investigation, which will undoubtedly teach
us much about subtle aspects of the Standard Model. Unfortunately, at
our present level of understanding we are not able to compute the
amplitudes from first principles, and are forced to make assumptions
about the non-perturbative QCD effects. Frequently these assumptions
concern factorization, i.e. whether matrix elements of operators in the
effective Hamiltonian which are products of two $V{-}A$ currents can be
written in terms of products of matrix elements of the currents, e.g.
\begin{equation}
\langle D^+\pi^-|\,(\bar d u)_L\,(\bar c b)_L|\,B^0\rangle
\stackrel{?}{=}\langle D^+|\,(\bar c b)_L |\,B^0\rangle
\ \langle \pi^-|\,(\bar d u)_L\,|\,0\rangle\ ,
\label{eq:factor}\end{equation}
where the first matrix element on the right hand side could be obtained
from semileptonic decays (see sec.~\ref{subsec:vcb}) and the second from
the known value of $f_\pi$. In some cases, the factorization
hypothesis can be motivated by the concept of \emph{colour
transparency}, that a light colour singlet meson, produced with large
energy at the weak decay vertex (and which is hence a small
colour-dipole), will travel a long way from the interaction region
before it grows to a size where it could interact
strongly~\cite{transparency,dugan}. It is possible that in some cases
factorization will give a reasonable first approximation and in many
others it will not be useful at all. Thus current analyses are limited
to a semi-quantitative level, and at this stage it is not possible to
endorse any of the approaches. I will not discuss these analyses in
these lectures, but to wish to point out the importance of this emerging
field, (see refs.~\cite{stech} and \cite{lp97} for more extensive
discussions).

\section{Conclusions}
\label{sec:concs}

\begin{figure}
\hbox to\hsize{\hfill
\begin{turn}{-90}
\epsfxsize0.5\hsize\epsffile{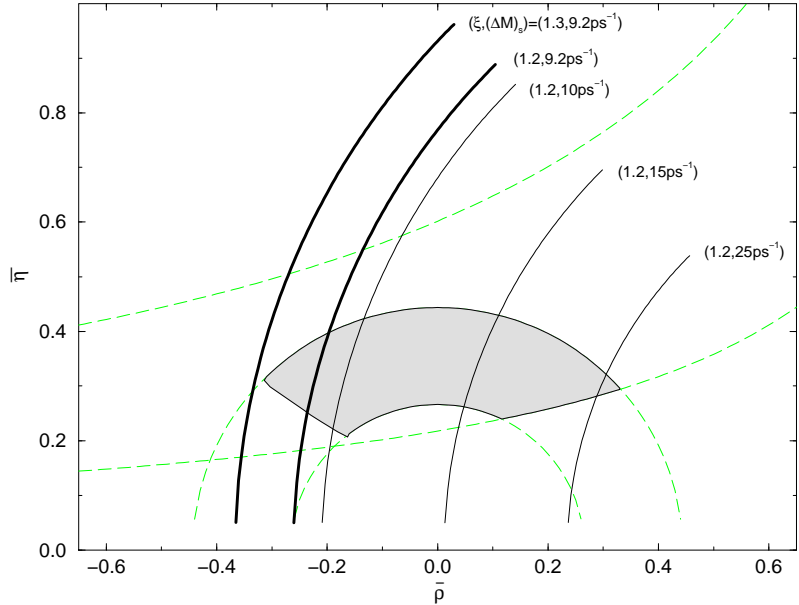}
\end{turn}\hfill}
\caption[]{The current status of our knowledge of the position 
of the vertex $A$ of the unitarity triangle as obtained by Buras
and Fleischer~\cite{bf}, (shaded area).}
\label{fig:rhoeta}
\end{figure}

In these lectures I have briefly reviewed the formalism required for the
study of the weak decays in the Standard Model of Particle Physics. 
From these studies we are attempting to measure the parameters of the
Standard Model, to understand the origin of $CP$-violation and to look
for signatures of physics beyond the Standard Model. I have explained
that the major theoretical difficulty in interpreting the wealth of
experimental data is our inability to control the non-perturbative
strong interaction effects to sufficient precision. Througout these
lectures I have illustrated the general discussion with applications to
various physical processes. The current status of the uncertainties in
determining the vertex $A$ of the unitarity triangle, from the analysis
of Buras and Fleischer is reproduced in fig.~\ref{fig:rhoeta}. The
circular arcs are obtained from information on $B-\bar B$ mixing, using
a modified analysis to that discussed in section~\ref{subsec:bbar}
above. In order to reduce the hadronic uncertainties, we can take the
ratio of the expressions for $\Delta M_{B_d}$ and $\Delta M_{B_s}$ in
eqs.~(\ref{eq:dmd}) and (\ref{eq:dms}), and, using the upper bounds on
$\Delta M_{B_d}$ (0.482~ps$^{-1}$) and $|V_{ts}/V_{cb}|$ (0.993), obtain
a bound for the quantity in eq.~(\ref{eq:vtdut}):
\begin{equation}
\frac{1}{\lambda}\frac{|V_{td}|}{|V_{cb}|}\,<\, 1.0\,\xi
\sqrt{\frac{10.2\,\textrm{ps}^{-1}}{\Delta M_{B_s}}}\ ,
\label{eq:rtmax}\end{equation}
where $\xi$ is defined in eq.~(\ref{eq:bresults}), where the lattice
result is also given. The curves in fig.~\ref{fig:rhoeta} represent
these bounds for several choices of $\xi$ and
$\Delta M_{B_s}$~\footnote{In fig.~\ref{fig:rhoeta} $\Delta M_{B_s}$ is
written as $\Delta M_s$.}.

I end these lectures by summarising the \emph{tourist guide} of Buras
and Fleischer~\cite{bf} to different processes, in which stars are
awarded to the processes depending on the level of theoretical
uncertainty.
\begin{enumerate}
\item[***] These are processes in which there are \emph{no} theoretical
uncertainties i.e. they are particularly small (less than 2\% say).
Among these is the mixing-induced $CP$-violating asymmetry in $B_d\to
J/\Psi K_S$ discussed in section~\ref{subsubsec:cpvneutral}, from which
we expect to determine the angle $\beta$ of the unitarity triangle.
Other quantities in this category which I have not had time to discuss
include the ratio of (inclusive) branching ratios $\textrm{Br}(B\to
X_d\nu\bar\nu)/\textrm{Br}(B\to X_s\nu\bar\nu)$ from which one would
obtain $|V_{td}|/|V_{ts}|$ and the decays $K_L\to\pi^0\nu\bar\nu$ (which
would give Im\,$\lambda_t$) and $K^+\to\pi^+\nu\bar\nu$ (which would
give $|V_{td}|$). For the kaon decays I have explained in
section~\ref{subsubsec:sinthetac}, that the  small mass difference
between the quarks in the $K$ and $\pi$ mesons implies that the hadronic
uncertainties are small.
\item[**\ ] The processes in this category have \emph{small} hadronic
uncertainties, of the order of 5-10\% say. They include the evaluation
of $V_{cb}$ discussed in sec.~\ref{subsec:vcb}, $\Delta M_d/\Delta M_s$
discussed in section~\ref{subsec:bbar} (from which we obtain
$|V_{td}|/|V_{ts}|$) and the mixing-induced $CP$-asymmetry in
$B_d\to\pi^+\pi^-$ (from which we would obtain the angle $\alpha$)
discussed in section~\ref{subsubsec:cpvneutral}. There is a large set of
other mixing-induced asymetries in this category, yielding, in
principle, all three angles. Being perhaps a little optimistic, the
determination of $|V_{ub}|/|V_{cb}|$ in inclusive decays, also belongs to
this category.

\item[*\ \ ] The processes in this category have \emph{smallish}
hadronic uncertainties, perhaps in the region of 15\% or so. Of the
quantities discussed in this course, $\Delta M_d$, $\Delta M_s$
(section~\ref{subsec:bbar}) and $\epsilon$
(section~\ref{subsubsec:epsilon}) belong in this category.

\item[.\ \ ] As we have seen in these lectures, there are also
important quantities for which the theoretical uncertainties are too
large for the experimental data to give accurate information about the
parameters of the Standard Model (and hence these quantities are not
ascribed any stars), but which nevertheless give interesting qualitative
information. These include $\epsilon^\prime/\epsilon$ (discussed in
section~\ref{subsubsec:deltai12}) and most $CP$-asymmetries in decays
of charged $B$'s. Non-leptonic decays of $B$-mesons in general belong
to this category.
\end{enumerate}

Of course it is to be hoped and expected that theoretical progress will
continue, and that the interesting physical quantities will gradually
acquire more stars. The theoretical community is working very hard to
achieve this.

\subsection*{Acknowledgments}

It is a pleasure to thank the students, lecturers and organisers of this
school for providing such a lively and stimulating atmosphere. It was
very thought-provoking and instructive to respond to the many excellent
questions from the students without being able to rely on an armoury of
theoretical formalism.

I warmly acknowledge my collaborators, particularly, Guido Martinelli
and Matthias Neubert, with whom I learned many of the topics discussed
in these lectures. I am grateful to Giulia de Divitiis and Luigi del
Debbio for their comments on the manuscript.

\def\warsaw{ICHEP96, 28th Int. Conf. on High Energy Physics, Warsaw,
Poland, 25--31 July 1996, edited by Z. Ajduk and A.K. Wroblewski,
World Scientific, Singapore (1997)}
\def\prd#1{Phys. Rev.  {\bf D#1}}
\def\prl#1{Phys. Rev. Lett. {\bf #1}}
\def\plb#1{Phys. Lett.  {\bf B#1}}
\def\npb#1{Nucl. Phys.  {\bf B#1}}
\def\npbps#1{Nucl. Phys. {\bf B (Proc. Suppl.)#1}}
\def\npaps#1{Nucl. Phys. {\bf A (Proc. Suppl.)#1}}
\def\zpc#1{Z. Phys {\bf C#1}}
\def\nima#1{Nucl. Instrum. Meth. {\bf A#1}}
\def\cmp#1{Commun. Math. Phys. {\bf #1}}
\def\physrep#1{Phys. Rep. {\bf #1}}
\def\edlat{Lattice 97, 15th Int. Symp. on Lattice Field Theory,
Edinburgh, Scotland, 1997}

\end{document}